\documentstyle[12pt]{article}

\newcommand \qed {\vrule height5pt width5pt}
\newcommand \weight {(|\epsilon| M)}

\newcommand \no {\noindent}
\newcommand \ra {\rightarrow}

\newcommand{\be}{\begin{equation}}
\newcommand{\ee}{\end{equation}}
\newcommand{\bea}{\begin{eqnarray}}
\newcommand{\eea}{\end{eqnarray}}
\newcommand \boundary {\partial}

\newcommand \f { {1 \over n !}}

\newcommand \s {\sigma}
\newcommand{\nn}{\hbox{nn}}

\newcommand \eps {\epsilon}
\newcommand \vareps {\varepsilon}

\newcommand{\ctrans} {{\bf T}}
\newcommand{\trans} {T}
\newcommand{\fun} {R}
\def\C{{\mathbf{C}}}
\newcommand{\sun} {U}
\newcommand{\hone} {H_1}
\newcommand{\xx} {{\cal X}}

\def\reff#1{(\ref{#1})}
\newtheorem{theorem}{Theorem}

\def \d {\triangle}
\def \L {\Lambda}
\def\zed{{\mathbf{Z}}}

\newcommand \fp {F} 

\begin{document}

\title{Instability of interfaces in the antiferromagnetic XXZ
chain at zero temperature}

\author{Nilanjana Datta 
\\Statistical Laboratory
\\Centre for Mathematical Sciences
\\University of Cambridge
\\Wilberforce Road, Cambridge CB30WB
\\ email: n.datta@statslab.cam.ac.uk
\\
\\Tom Kennedy
\\Department of Mathematics
\\University of Arizona
\\Tucson, AZ 85721
\\ email: tgk@math.arizona.edu
\bigskip
}

\maketitle

\begin{abstract}
For the antiferromagnetic, highly anisotropic XZ and XXZ quantum spin 
chains, we impose periodic boundary conditions on chains with an odd
number of sites to force an interface (or kink) into the chain. 
We prove that the energy of the interface depends on the momentum of the
state. This shows that at zero temperature the interface in such chains 
is not stable. This is in contrast to the ferromagnetic XXZ chain for
which the existence of localized interface ground states has been proven
for any amount of anisotropy in the Ising-like regime. 
\end{abstract}

\newpage

\section{Introduction}

Interfaces or domain walls in classical spin systems have been the subject
of mathematical study for several decades. 
Dobrushin proved \cite{Dob} that in the three-dimensional Ising
model at low temperatures, under suitable (Dobrushin) boundary conditions,
there is a stable interface orthogonal to the $001$--direction.
These boundary conditions hence yield a non-translation invariant 
Gibbs state at low temperatures. However, Gallavotti proved
\cite{Gal} that the two--dimensional model
shows a very different behavior; thermal
fluctuations destabilize the interface and the corresponding Gibbs
state is translation invariant.

Interfaces in quantum-mechanical systems can exhibit a much richer 
and more complex behavior than 
their classical counterparts. A review of some of this behavior may be 
found in \cite{nac}. For example, quantum fluctuations may
lift a classical degeneracy and, in doing so, stabilize an 
interface (against thermal fluctuations)
that is unstable in the corresponding classical system. 
Such a stabilization
is an example of the phenomenon of ground state selection \cite{Hen}. 
It is expected to occur for the 
$111$--(or diagonal) interface in the three-dimensional 
ferromagnetic, anisotropic XXZ model [see e.g. \cite{BCNS1, BCNS2}], and has been proved to occur for the
$111$-interface in the three-dimensional Falicov-Kimball model \cite{DMN}.
These models can be viewed as quantum perturbations of the classical
Ising model. In contrast to these quantum--mechanical models,
the diagonal interface in the three--dimensional classical Ising model 
is expected to be unstable at non--zero temperatures. This is due 
to the massive degeneracy of the zero-temperature configurations 
compatible with the boundary conditions which favor
such an interface [see \cite{ken}]. 

Another interesting feature of interfaces in quantum-mechanical systems is
the diverse nature of the low--lying excitations above the interface
ground states for different models and for different orientations
of the interface. For example, there are gapless excitations above the 
conjectured diagonal interface states in the spin-$1/2$ 
ferromagnetic, anisotropic, XXZ model. These excitations were 
described in the two-dimensional case by Koma and Nachtergaele 
\cite{BCNS2,KNS1,KN3},
and proved to exist in all dimensions greater than one by Matsui 
\cite{Mat2}. In contrast, it is expected that there is a gap in 
the spectrum above a ground state that describes an interface perpendicular
to a coordinate direction.

For quantum-mechanical systems, the stability of an interface is a nontrivial 
question even in the ground state, since quantum fluctuations
can destabilize the interface at zero temperature. In this case
quantum fluctuations play a role analogous to that of thermal
fluctuations in classical systems. 
In one dimension we expect interface states to be unstable 
for generic Hamiltonians. However, there are notable
exceptions, e.g. the anisotropic ferromagnetic XXZ chain.
In addition  to its two ferromagnetically ordered,
translation invariant ground states, this model has ground states
corresponding to an interface between two domains
of opposite magnetization. The stability of this interface 
was proved independently by Alcaraz, Salinas and Wreszinski \cite{ASW} and
Gottstein and Werner \cite{GW}. This stability is a direct 
consequence of the 
conservation of the total $z$-component of the spin. There are no terms
in the Hamiltonian that can simply move the interface across
one lattice spacing. To conserve the spin, one must at the same time 
create a new excitation
in the chain, thus raising the energy of the state. More precisely, it
was proved in \cite{ASW, GW} that, under suitable boundary conditions,
there exists a family of interface ground states which describe a 
localized domain wall. The localization length 
depends on the anisotropy of the model
and diverges in the limit of the isotropic model. Alternative proofs
of the stability of this interface were given in \cite{BCN}, 
by using the path integral representation of interface states,
and in \cite{bach}, by employing the principle of
exponential localization \cite{FL}. The above results show that in the
spin-$1/2$ ferromagnetic, anisotropic XXZ model, an arbitrarily small amount of anisotropy
is sufficient to stabilize the interface against quantum fluctuations.

Quantum perturbations do not always have the drastic effect of
either stabilizing an unstable classical interface or destabilizing
an interface at zero temperature. There exist quantum lattice models
which are quantum perturbations of suitable classical systems such
that an interface in the classical system remains essentially
unchanged under the quantum perturbations. 
For example, if we add a 
quantum perturbation to the three dimensional Ising model, then 
the so-called Dobrushin condition induces a stable 
interface in the system, in the sense that there is a low temperature 
non--translation invariant Gibbs state describing an asymptotically
horizontal interface. This was proved in a more general setting by 
Borgs, Chayes, and Fr\"ohlich \cite{BCF} for systems in dimensions
$d \ge 3$, by using a quantum version of the Pirogov Sinai theory
\cite{bku, dff}. One expects that adding 
a quantum perturbation to the two-dimensional Ising model at low temperatures
will not stabilize the $10$-interface in this model 
but we are not aware of any proof of this. 

In this paper we consider the stability of the interface states in the 
anisotropic, antiferromagnetic(AF) XXZ and XZ models at zero temperature. 
We prove that in these models the interface is not stable in
one dimension. 
We study the question of stability 
by analyzing the dispersion relation for the energy of the interface, i.e.,
its energy as a function of its momentum. 
For the AF models we can force an interface into the system by 
imposing periodic boundary conditions on a chain with an odd number of 
sites. We can study the energy of the interface by comparing the energies
for chains with an even and odd number of sites. 
The AF Hamiltonians that we consider are invariant both 
under lattice translations and global spin flips. The combined symmetry of 
translating by one lattice spacing and then performing a global spin flip,
which we denote by ${\widetilde{\ctrans}}$, is a useful symmetry 
for studying the interfaces since it leaves the N\'eel states invariant. 
We refer to the eigenvalue of this symmetry operator 
as a ``generalized momentum.'' 
We study the difference between the lowest energy of an eigenstate 
with generalized momentum $k$ for a chain with an 
odd number of sites and that with an even number of sites.
We take this difference to be the definition of the dispersion relation
for the interface.
If the interface is stable, then there should be an eigenstate 
$|\Psi\rangle$ of the Hamiltonian (for a chain with an odd number of
sites) which has some localized structure. 
So the states ${\widetilde{\ctrans}}^l |\Psi\rangle$ should be linearly independent. 
By taking linear combinations of these states, 
\be 
|\Phi_k\rangle = \sum_l e^{ikl}{\widetilde{\ctrans}}^l |\Psi\rangle,
\ee
we can form eigenstates of the Hamiltonian with different generalized momenta. 
Since ${\widetilde{\ctrans}}$ commutes with the Hamiltonian, 
these states all have 
the same energy. Thus the dispersion relation is independent of 
the generalized momentum if there is a stable interface.
We prove that in the infinite volume limit the 
dispersion relation for the AF chain depends on the 
generalized momentum, and so the chain does not admit ground states 
that correspond to a stable interface. In contrast, for the
anisotropic, ferromagnetic XXZ chain, we prove that the dispersion
relation is ``flat'' (i.e., $k$--independent) in the infinite length limit. 
This provides another approach to studying the stability of the interface 
in this model at zero temperature to complement the approaches of 
\cite{ASW, GW, BCN, bach}. 

The XZ chain is exactly solvable, and Araki and Matsui used this to prove
the absence of non-translationally invariant infinite volume ground states
\cite {AM}. 
This shows the interface is unstable in this model since infinite volume
ground states containing an interface would be non-translationally 
invariant. The XXZ model is also exactly solvable, so one might be able 
to use this solvability to study the dispersion relation we study.
We emphasize, however, that in our approach we do not use the exact 
solvability of either of these models. 
The techniques that we use to study the interface are based on a novel 
approach to the analysis of ground states of quantum spin systems, introduced
by Kirkwood and Thomas \cite{KT}. 
They considered spin--$1/2$ models, but their approach was applied to some
higher spin models by Matsui \cite{Mat1}. Their method originally required a 
Perron-Frobenius condition on the Hamiltonian. We removed this 
condition and simplified the proof of convergence of the expansion
in \cite{DK}. Although we restrict our attention to the XZ and XXZ 
models in this paper, we expect the methods and results to 
be applicable to a much broader class of models. 

The paper is organized as follows: To keep the paper self--contained,
we first give a summary of our version of the Kirkwood--Thomas approach
(as developed in \cite{DK}) by using it to study the ground state
of the AF anisotropic XZ Hamiltonian. This is done
in Section \ref{groundxz} for a $d$--dimensional lattice under periodic
boundary conditions. The results of this section, for the case $d=1$, are 
used later in our analysis of interface states in the AF anisotropic XZ
chain. If the number of sites $N$ in such a chain is even then the 
ground state does not have an interface. However, if $N$ is odd then
the  periodic boundary conditions force an interface in the chain. The 
latter situation is studied in Section \ref{intfxz}. We prove that the 
dispersion relation for the energy of the interface depends non--trivially
on the generalized momentum $k$ even in the limit $N \rightarrow \infty$.
This allows us to conclude that the ground state of the 
AF anisotropic XZ chain does not have a stable interface. In Section \ref{af}
we prove a similar result for the AF anisotropic XXZ chain. In contrast, 
in Section 5, we prove that for the corresponding 
ferromagnetic model the energy of the interface does not depend on 
$k$ in the limit $N \rightarrow \infty$. 

\section{XZ ground state : the Kirkwood--Thomas approach}
\label{groundxz}
We consider the following antiferromagnetic 
Hamiltonian defined on a finite lattice $\Lambda \subset
\zed^d$
\be
{\widetilde{H}}
= \sum_{\langle ij\rangle \subset \Lambda} \s^z_i \s^z_j 
+ \eps \sum_{\langle ij\rangle \subset \Lambda} \s^x_i \s^x_j, 
\label{hamorig}
\ee
where the sums are over all nearest neighbor pairs (denoted 
by $\langle ij\rangle $) in $\Lambda$. We impose periodic boundary conditions
and assume that $\Lambda$ has an even number of sites in each 
coordinate direction. The Hamiltonian ${\widetilde{H}}$ acts on
the Hilbert space ${\cal H}_{\Lambda}= (\C^2)^{\otimes |\Lambda|}$,
where $|\Lambda|$ denotes the number of sites in the lattice $\Lambda$.
The Hamiltonian and most of the 
quantities that follow depend on the volume $\Lambda$. However, 
for notational simplicity, we often suppress this explicit 
dependence. 
The above Hamiltonian commutes with the global spin flip operator given by
\be 
{\widetilde{P}}
= \prod_{i \in \L} \s^x_i.
\label{ptilde}
\ee

The above form of the Hamiltonian seems natural for perturbation
theory in $\eps$ since the $\eps=0$ Hamiltonian is diagonal. 
However, following Kirkwood and Thomas, we 
study a unitarily equivalent Hamiltonian obtained by 
a rotation about the $Y$--axis in spin space caused by the 
operator
\be R = \exp\left(i\frac{\pi}{4} \sum_{j \in \Lambda} \sigma_j^y\right).
\label{rotop}
\ee
Hence, 
$$R\sigma_i^x R^{-1} = \sigma_i^z \,\,; \quad 
R \sigma_i^z R^{-1} = - \sigma_i^x, $$
and therefore 
\be
\fun {\widetilde{H}} \fun^{-1} = \sum_{\langle ij\rangle  \subset \Lambda} \s^x_i \s^x_j 
+ \eps \sum_{\langle ij\rangle  \subset \Lambda} \s^z_i \s^z_j .
\label{eqb}
\ee
The global spin flip operator transforms into
\be 
\fun {\widetilde{P}} \fun^{-1} = \prod_{i \in \L} \s^z_i .
\label{proj}
\ee
Finally, we perform a unitary transformation to change the $\epsilon=0$
Hamiltonian from antiferromagnetic to ferromagnetic. Define
\be
\sun = \prod_{j \in \Lambda \atop{j \, odd}} \sigma^z_j
\label{sundef}
\ee
where $j \, odd$ means that the sum of the components of $j$ is odd. 
Since $\Lambda$ has an even number of sites in each coordinate direction,
the transformed Hamiltonian, $H$, is given by
\be
H= \sun \fun {\widetilde{H}} \fun^{-1} \sun^{-1} = 
  - \sum_{\langle ij\rangle  \subset \Lambda} \s^x_i \s^x_j 
+ \eps \sum_{\langle ij\rangle  \subset \Lambda} \s^z_i \s^z_j .
\label{eqbu}
\ee
Since $[H,P]=0$, the state space of the Hamiltonian $H$ can be   
decomposed into two subspaces corresponding to the eigenvalues $+1$
and $-1$ of $P$. We refer to these two subspaces as the even and 
odd sectors respectively. The transformed global spin flip operator, 
$\fun {\widetilde{P}} \fun^{-1}$, remains unchanged under the 
action of the unitary operator $\sun$: 
\be
P= \sun \fun {\widetilde{P}} \fun^{-1} \sun^{-1} =  \prod_{i \in \L} \s^z_i.
\label{proju}
\ee
We emphasize that eq.\reff{eqbu} is not true if $\Lambda$ has an odd number
of sites in any lattice direction. This fact plays a key role in our 
study of interfaces in the one dimensional case 
[see e.g. Section \ref{intfxz}].

Let us introduce some definitions and notations. 
A classical spin configuration on the lattice is defined to
be an assignment
of a $+1$ or a $-1$ to each site in the lattice. Hence, for each
$i \in \L$, $\s_i = \pm 1$. We will abbreviate the classical
spin configuration $\{\s_i\}_{i \in \L}$ by $\s$ .
For each such $\s$ we let $|\s\rangle$ be the 
state in the Hilbert space, ${\cal H}_{\Lambda}$, which is the tensor product 
of a spin--up state at each site with $\s_i=+1$ and a spin--down state
at each site with $\s_i=-1$. 
Thus $|\s\rangle$ is an  eigenstate of all the $\s^z_i$ with 
$\s^z_i |\s\rangle = \s_i |\s\rangle$. The states  $|\s\rangle$
form a complete orthonormal basis of ${\cal H}_{\Lambda}$.
Any state $|\Psi\rangle$ can be written in terms of this basis:
\be
|\Psi\rangle = \sum_\s \psi(\s) |\s\rangle 
\label{eqa}
\ee
where $\psi(\s)$ is a complex-valued function on the spin configurations
$\s$.
For a single site, the vectors $\left(|+1\rangle + |-1\rangle\right)$ and 
$\left(|+1\rangle - |-1\rangle\right)$ are the eigenstates of 
$\s^x$ with eigenvalues $+1$ and $-1$, respectively. 
Thus the (unnormalized) 
ground states of the Hamiltonian, $H$, [\reff{eqbu}] for $\eps=0$ are
given by \reff{eqa} with $\psi(\s)=1$ and 
$\psi(\s)= \prod_{i \in \Lambda} \s_i$. 
We define 
\be
\s(X) = \prod_{i \in X} \s_i
\ee 
and use the convention that $\s(\emptyset)=1$.
Note that $\sigma (\Lambda)$ 
is equal to $+1 (-1)$ in the even (odd) sector. 

\medskip

In the Kirkwood--Thomas method one expands the ground state 
with respect to the basis $\{|\s\rangle\}$, as 
in eq. \reff{eqa}, and writes $\psi(\s)$ in the form 
\be
\psi(\s)= \exp[- {1 \over 2} \sum_X g(X) \s(X)] 
\label{eqc}
\ee
for some real $g(X)$. 
As in \cite{DK}, we justify the above exponential form of
$\psi(\s)$ by a two--step procedure: 
First, we consider \reff{eqc} to be an ansatz and prove that
it satisfies the Schr\"odinger equation. This ensures that there
is an eigenstate of the form \reff{eqc}. Next we give
an argument to show that this eigenstate 
must in fact be the ground state.

Consider the Schr\"odinger equation 
\be
H \Psi =  E_0 \Psi 
\label{eqbb}
\ee
The operator $\s^z_i \s^z_j$ is diagonal in the chosen basis, so 
\be
\s^z_i \s^z_j \sum_\s \psi(\s) |\s\rangle  
= \sum_\s \s_i \s_j \psi(\s) |\s\rangle  .
\ee
The operator $\s^x_i \s^x_j$ just flips the spins at sites $i$ and $j$, i.e.,
$\s^x_i \s^x_j |\s\rangle  = |\s^{(ij)}\rangle $, where $\s^{(ij)}$ is the spin configuration
$\s$ but with $\s_i$ replaced by $-\s_i$ and  $\s_j$ replaced by $-\s_j$.
Hence
\be
\s^x_i \s^x_j \sum_\s \psi(\s) |\s\rangle  
= \sum_\s \psi(\s) |\s^{(ij)}\rangle 
= \sum_\s \psi(\s^{(ij)}) |\s\rangle  .
\ee
The last equality follows by a change of variables in the sum.

We now see that if we 
use \reff{eqa} in the Schr\"odinger equation \reff{eqbb} and pick out the 
coefficient of $|\s\rangle $, then for each spin configuration $\s$ we have
\be - \sum_{\langle ij\rangle } \psi(\s^{(ij)}) + \eps \sum_{\langle ij\rangle } \, \s_i \s_j \psi(\s)
= E_0 \psi(\s).
\label{one5}
\ee
Henceforth, the condition $\langle ij\rangle \subset \Lambda$ 
will be implicit in 
all our sums on $\langle ij\rangle $.
Dividing both sides of \reff{one5} by $\psi(\s)$ we have 
\be
- \sum_{\langle ij\rangle } { \psi(\s^{(ij)}) \over \psi(\s)}
 + \eps \sum_{\langle ij\rangle } \, \s_i \s_j = E_0.
\ee
Now $\s^{(ij)}(X)$ is $\s(X)$ when both of $i$ and $j$ are in $X$,
and when both of them are not in $X$. If exactly one of $i$ and $j$ is 
in $X$, then $\s^{(ij)}(X)$ is $-\s^{(ij)}(X)$.
We will let $\boundary X$ denote the set of nearest neighbor 
bonds which connect a
site in $X$ with a site not in $X$. (Henceforth, we will always use the
word bond to denote a nearest neighbor bond.) 
Then the condition that exactly
one of $i$ and $j$ belongs to $X$ may be written as $\langle ij\rangle  \in \boundary X$. 
We will often abbreviate this condition as $X : \langle ij\rangle $.
Thus
\be
\psi(\s^{(ij)}) =\exp[- {1 \over 2} \sum_X g(X) \s(X)
+ \sum_{X: \langle ij\rangle  } g(X) \s(X)]
\ee
and so the Schr\"odinger  equation is now
\be
- \sum_{\langle ij\rangle } \exp[\sum_{X: \langle ij\rangle } g(X) \s(X)]
+ \eps \sum_{\langle ij\rangle } \, \s_i \s_j = E_0. 
\label{eqd}
\ee
As in \cite{DK}, we refer to this equation as the Kirkwood-Thomas equation.

We expand the exponential in a power series. The contribution from the linear 
term may be rewritten as 
\be \sum_{\langle ij\rangle } \, \, \sum_{X: \langle ij\rangle } g(X) \s(X)
= \sum_X |\boundary X| g(X) \s(X)
\ee
where $| \boundary X |$ 
is the number of  bonds in $\boundary X$, i.e., the number 
of  bonds that connect a site in $X$ with a site not in $X$. 
Hence the Kirkwood Thomas equation becomes
\bea
\sum_X |\boundary X| g(X) \s(X) + E_0 + d |\L|
&=& 
 - \sum_{\langle ij\rangle } \sum_{n=2}^\infty \, \f 
\sum_{X_1,X_2,\cdots,X_n: \langle ij\rangle }
\prod_{k=1}^n  g(X_k) \s(X_k)
+ \eps \sum_{\langle ij\rangle } \, \s_i \s_j.\nonumber\\
\label{kt}
\eea
Here $d |\L|$ is the number of  bonds in the lattice.  

Since $\s_i^2=1$, $\s(X) \s(Y) = \s(X \d Y)$ where the
symmetric difference $X \d Y$ of $X$ and $Y$ is defined by
$X \d Y = X \cup Y \setminus (X \cap Y)$. 
Thus
$\prod_{k=1}^n \s(X_k) = \s(X_1 \d \cdots \d X_n)$. 
If we equate the coefficient of $\s (X)$ on both sides of eq. \reff{kt},
we obtain, for $X \ne \emptyset$, 
\be
 g(X) 
= \frac{1}{|\boundary X|} \left[ 
 - \sum_{\langle ij\rangle } \sum_{n=2}^\infty \, \f 
\sum_{X_1,X_2,\cdots,X_n: \langle ij\rangle , 
\atop{X_1 \d \cdots \d X_n=X} } 
g(X_1) g(X_2) \cdots g(X_n) 
\, + \, \eps \, 1_{\nn}(X)\right]. 
\label{fixedpt}
\ee
where $1_{\nn}(X)$ is $1$ if $X$ consists of two nearest neighbor 
sites and is $0$ otherwise.

If $X=\L$, then $\boundary X = \emptyset$. So the coefficient of 
$g(\L)$ on the LHS of equation \reff{kt} is zero. 
This looks like a fatal problem since the RHS of the 
equation will contain a multiple of $\s(\L)$. 
We solve this problem by 
exploiting the decomposition of the state space 
into even and odd sectors (as in \cite{KT}). We look for eigenstates
of the form 
\be
|\Psi_e\rangle = \sum_{\s: {\rm{even}}} \psi(\s) |\s\rangle  
\label{eqeven}
\ee
and 
\be
|\Psi_o\rangle = \sum_{\s:{\rm{odd}}} \psi(\s) |\s\rangle  
\label{eqodd}
\ee
where the sums are only over configurations $\s$ for which 
the number of sites $i$ with 
$\s_i=-1$ is even or odd, respectively. 
(Equivalently, $\sigma (\Lambda) := \prod_{i \in \L} \s_i=+1$, or $-1$.)
The Schr\"odinger equation is still equivalent to \reff{eqd}, but 
now to find an eigenstate in the even (respectively, odd) sector, this
equation need only hold for $\s$ with  
$\prod_{i \in \L} \s_i= +1$ (respectively, $-1$).
Thus the terms on the RHS of \reff{kt} which 
contain $\s(\L)$ may be included in the equation for $X=\emptyset$. 
So for $X=\emptyset$, we obtain the equation
\bea
E_{\pm} + d |\L| && =  - \sum_{\langle ij\rangle } \sum_{n=2}^\infty \, \f 
\sum_{X_1,X_2,\cdots,X_n: \langle ij\rangle ,
\atop{X_1 \d \cdots \d X_n= \emptyset} } 
g(X_1) g(X_2) \cdots g(X_n) \nonumber\\
&& \mp \sum_{\langle ij\rangle } \sum_{n=2}^\infty \, \f 
\sum_{X_1,X_2,\cdots,X_n: \langle ij\rangle ,
\atop{X_1 \d \cdots \d X_n= \L } } 
g(X_1) g(X_2) \cdots g(X_n).
\label{energypm}
\eea
Here and henceforth, the upper (lower) sign corresponds to the 
even (odd) sector.
We have replaced $E_0$ by $E_\pm$ since the 
eigenvectors in the even and odd sectors have different eigenvalues. 
We will see later that the difference between the two eigenvalues is 
exponentially small in the number of sites in the lattice
$\Lambda$. 
Note that eq. \reff{eqd} for the two sectors can be combined into the
single equation
\be
  - \sum_{j=1}^N \exp(\sum_{Y : j} g(Y) \s(Y))  
 + \epsilon \sum_{j=1}^N \, \s_j \s_{j+1} = {E_+ + E_- \over 2} 
 + {E_+ - E_- \over 2} \s(\Lambda).
\label{eqdd} 
\ee

We let $g$ denote the collection of coefficients 
$\{g(X): X \subset \Lambda, X \ne \emptyset, \, X \ne \L \}$,
and think of eq.\reff{fixedpt} as a fixed point equation, $g=F(g)$.  
We define a norm by 
\be
||g||= \sum_{X: b} \, |g(X)| \, |\boundary X| \, \weight^{-w(X)}, 
\label{normg}
\ee
where $b$ is a nearest neighbor 
bond and $w(X)$ is defined as follows: We consider two  bonds to 
be ``connected'' if they share an endpoint or if the distance between
them is 1. We consider a set of  bonds to be ``connected'' if we can 
get from one bond in the set to any other bond in the set by going through a 
sequence of connected  bonds in the set.
Then $w(X)$ is the cardinality of the smallest set of  bonds
which contains $X$ and is ``connected.'' Note that the symmetries
of the lattice imply that the norm $||g||$ 
does not depend on the choice of $b$.

\begin{theorem}
There exists a constant $M>0$ which depends only on the number of 
dimensions of the lattice, such that if $|\eps| M \le 1$,
then the fixed point equation \reff{fixedpt} has a solution $g$, 
and $||g|| \le \delta $ for some constant $\delta$ which depends 
only on the lattice.
\label{theo1}
\end{theorem}

\no {\bf Proof:} We will prove that $F$ is a contraction on 
a small ball about the origin, and that it maps this 
ball back into itself. The contraction mapping theorem
will then imply that $F$ has a fixed point in this ball. 
For the sake of concreteness, we prove it is a contraction 
with constant $1/2$, but there is nothing special about the choice of 
$1/2$. 

Define 
\be
\delta = {4(2d -1) \over M}
\label{del}
\ee
We will show that 
\be
||\fp(g)-\fp(g')|| \le \frac{1}{2} ||g - g'|| \quad {\hbox{for}} \quad
||g||, ||g'|| \le \delta, 
\label{boundone}
\ee
and 
\be
||\fp(g)|| \le \delta \quad {\hbox{for}} \quad
||g|| \le \delta .
\label{boundtwo}
\ee

The proof of \reff{boundone} proceeds as follows:
Fix a  bond $b$ to use in the definition of $||F(g)-F(g')||$. Then
\be
||F(g)-F(g')|| \le 
\sum_{\langle ij\rangle } \sum_{n=2}^\infty {1 \over n!}
\sum_{X_1, \cdots, X_n : \langle ij\rangle , b \in \boundary \Delta} 
|g(X_1) \cdots g(X_n) - g'(X_1) \cdots g'(X_n)| \weight^{-w(\Delta)}, 
\label{return}
\ee
where $\Delta = X_1 \d \ldots \d X_n$. 
If $b \in \boundary \Delta$, then $b$ is in at least one $\boundary X_k$. 
Using the symmetry under permutations of the $X_k$, we can 
take $b \in \boundary X_1$ at the cost of a factor of $n$. 
We claim that if $\langle ij\rangle \in \boundary X_k$ 
for $k=1,2,\cdots,n$, then 
\be
w(X_1 \d \cdots \d X_n) \le \sum_{k=1}^n w(X_k) .
\label{wtri}
\ee
To prove the claim, 
for $k=1,2,\cdots,n$, let $C_k$ be sets of  bonds such that 
$X_k \subset C_k$, $|C_k|=w(X_k)$ and $C_k$ is connected in the 
sense used to define $w(X_k)$ [see discussion after \reff{normg}]. 
Define $C=\cup_{k=1}^n C_k$. 
Since $X_k$ contains exactly one of the sites $i$ and $j$, 
$C_k$ contains at least
one of the sites $i$ and $j$. 
Since $C_1, \ldots, C_n$ are connected 
this implies that $C$ is connected.
Clearly, $X_1 \d \cdots \d X_n \subset C$.
So 
\be w(X_1 \d \cdots \d X_n) \le |C| \le \sum_{k=1}^n |C_k|
= \sum_{k=1}^n w(X_k),
\label{mtri}
\ee
which proves the claim \reff{wtri}.

Using 
\be
|\prod_{k=1}^n g(X_k) - \prod_{k=1}^n g'(X_k)|
\le
\sum_{k=1}^n \prod_{i=1}^{k-1} |g(X_i)| \, |g(X_k)-g'(X_k)|
\prod_{i=k+1}^n |g(X_i)| 
\ee
we have 
\bea
||F(g) - F(g')|| 
&\le & 
\sum_{n=2}^\infty {1 \over (n-1)!}
\sum_{X_1: b } \, \,
\sum_{\langle ij\rangle  \in \boundary X_1} \, \,
\sum_{X_2, \ldots, X_n : \langle ij\rangle } \, \, \sum_{k=1}^n 
\Bigl(\prod_{i=1}^{k-1} |g(X_i)| \weight^{-w(X_i)}\Bigr) 
\nonumber\\
& & \quad \quad \times |g(X_k)-g'(X_k)| \weight^{-w(X_k)} \Bigl(\prod_{i=k+1}^{n} |g'(X_i)| \weight^{-w(X_i)}\Bigr)
\nonumber\\
&\le &  \sum_{n=2}^\infty {1 \over (n-1)!} \,\, || g - g'|| \,\,
\sum_{k=1}^n ||g||^{k-1} \, ||g'||^{n-k} \nonumber\\
& \le & K ||g-g'||,\nonumber\\
\label{g11}
\eea
where
\be
K = \sum_{n=2}^\infty {n \over (n-1)!} \delta^{n-1}
= e^{\delta} - 1 + \delta \, e^{\delta},
\label{kg}
\ee
and we have used the fact that both $||g||$ and $||g'||$ are bounded
by $\delta$. 
By choosing $\delta$ to be sufficiently small we obtain $K \le 1/2$.

To prove \reff{boundtwo}, we use \reff{boundone} with $g'=0$. 
{From} \reff{fixedpt} it follows that
\bea
||\fp(0)|| &\le& \sum_{X : b} \,\, \eps \,\,
1_{\nn} (X) \, \weight^{-w(X)}\nonumber\\
& \le & 2(2d -1) \eps \weight^{-1}
 \le  2(2d -1) M^{-1} 
= {\delta \over 2}.
\nonumber\\
\label{zero}
\eea
Hence,
\bea
||\fp(g)|| &\le& ||\fp(g)-\fp(0)|| + ||\fp(0)||
\le {1 \over 2} ||g|| + {1 \over 2} \delta
\le \delta
\eea
\qed

Eq. \reff{energypm} may be used to study the difference between the 
ground state energies in the odd and even sectors. It is straightforward
to show that 
\be 
|E_{-} - E_+| \le c \weight^{w(\Lambda)}\, d|\Lambda|,\, 
\ee
where the constant $c$ depends on $||g||$.
Since $w(\Lambda)=|\Lambda|/2$, the difference between these two 
eigenvalues is exponentially small in the number of sites in the lattice. 

We conclude this section by showing that the eigenstates we have constructed
in the even and odd sectors are indeed the lowest eigenstates in these
sectors. The argument is similar to that in \cite{DK}, but some small 
modifications are needed to take account of the decomposition into even and 
odd sectors. 
We know our eigenstates are the lowest in their sectors 
when $\epsilon=0$. Since we have a finite
lattice, our eigenvalue problem is finite dimensional. So in each 
sector, our eigenstate will remain the lowest eigenstate provided its 
eigenvalue does not cross another eigenvalue associated with that sector, 
i.e., provided the eigensubspace in the sector associated with our 
eigenvalue continues to be one-dimensional. Hence, if we show that there 
exists an $\epsilon_0>0$ such that our eigenfunction is non--degenerate
for all $\epsilon$ with $|\epsilon| <\epsilon_0$, then it would follow that
our eigenfunction is the ground state for all such $\epsilon$.

Suppose that there is a value of $\epsilon$ for which there is another 
eigenvector $|\Psi^\prime_e\rangle$ with the same 
eigenvalue as $|\Psi_e\rangle$. 
(The argument in the case of the odd sector is identical.)
Define $\psi^\prime(\s)$ for even $\s$ by 
\be
|\Psi^\prime_e\rangle = \sum_{\s: {\rm{even}}} \psi^\prime(\s) |\s\rangle,  
\ee
and let $\psi^\prime(\s)=0$ for odd $\s$. 
Now consider $\psi(\s)+\alpha \psi^\prime(\s)$ where $\alpha$ is a small real 
number and $\psi(\s)$ is defined through \reff{eqeven}. As $\alpha \ra 0$, this converges to $\psi(\s)$ 
for each $\s$. There are only finitely many values of $\s$, so for small 
enough $\alpha$, this function is always positive (since $\psi(\s)>0 \,\forall
\, \s$). So it can be written 
as $\exp[-{1 \over 2} \sum_X g_\alpha(X) \s(X)]$. Moreover, as 
$\alpha \ra 0$, $g_\alpha(\s) \ra g(\s)$ for each $\s$, and by construction
$g_\alpha$ satisfies the fixed point equation. So for sufficiently small
$\alpha$, $g_\alpha$ is a solution of the fixed point equation  
which is inside the ball in which we know the fixed point equation
has a unique solution. This contradiction completes the argument. 

\section{Interfaces in the Antiferromagnetic XZ chain} 
\label{intfxz}

In this section we consider the model of the previous section in
one dimension. So $\Lambda=\{1,\cdots,N\}$, and 
\be
{\widetilde{H}} = \sum_{j=1}^N \s^z_j \s^z_{j+1} 
+ \epsilon \sum_{j=1}^N \s^x_j \s^x_{j+1} .
\label{hamorg}
\ee
The indices should be taken to be periodic, e.g., 
$\s^x_{N+1}$ means $\s^x_1$.  
When $N$ is even, we have as before 
\be
H= \sun \fun {\widetilde{H}} \fun^{-1} \sun^{-1} = 
  - \sum_{j=1}^N \s^x_j \s^x_{j+1} 
+ \eps \sum_{j=1}^N \s^z_j \s^z_{j+1}, 
\label{four0}
\ee
and the ground state may be constructed as in the previous section.
If $N$ is odd, then the periodic boundary conditions force an interface
into the antiferromagnetic chain. In this case we have 
\be
H= \sun \fun {\widetilde{H}} \fun^{-1} \sun^{-1} = 
  - \sum_{j=1}^N J_j \s^x_j \s^x_{j+1} + \eps \sum_{j=1}^N \s^z_j \s^z_{j+1} , 
\label{four1}
\ee
where the coupling $J_j$ is $+1$ except when $j=N$, in which case it is $-1$.  
So the $\epsilon=0$ Hamiltonian has ferromagnetic couplings for all the  bonds
except the bond between the sites $1$ and $N$. 

As before the Hamiltonian ${\widetilde{H}}$ commutes with the global 
spin flip operator ${\widetilde{P}}$ [\reff{ptilde}]. It also commutes
with the translation operator $\trans$ defined by 
\be 
\trans \s^\alpha_i \trans^{-1}= \s^\alpha_{i+1}, \quad \alpha=x,y,z.
\ee
When $\epsilon=0$ and $N$ is even, the ground states 
of the Hamiltonian $\widetilde{H}$ [\reff{hamorg}] are the two N\'eel states.
These states are not invariant under translation. However, if we 
translate and then perform the global spin flip, the N\'eel states 
remain unchanged. So if we define 
\be 
{\widetilde{\ctrans}}={\widetilde{P}} \trans
\ee
then ${\widetilde{\ctrans}}$ commutes with ${\widetilde{H}}$ and leaves the 
N\'eel states invariant. This combined symmetry of the Hamiltonian will be the 
most useful one in our study of interface states, since its action on an
interface is to simply translate the interface by one site.
Let 
\be
\ctrans= \sun \fun {\widetilde{\ctrans}} \fun^{-1} \sun^{-1}
\label{tee}
\ee 
be this combined symmetry after our unitary transformations.
Simple calculations show that when $N$ is even,
$ \ctrans$ is equal to the pure translation operator $\trans$. 
However, for odd values of $N$ we find that 
\be
\ctrans = \s^z_1 \trans.
\label{genT}
\ee
In words, $\ctrans$ translates by one lattice spacing 
and rotates the spin at the site $i=1$. 
We can refer to it as a generalized translation operator.
Throughout this section we will assume $N$ to be odd. 

Since $H$ and $\ctrans$ commute, we choose the eigenfunctions of 
$H$ to be eigenfunctions of $\ctrans$ as well.
So they can be labeled by an index $k$, 
where $k$ can be regarded as the generalized ``momentum'', i.e.,
\be
\ctrans \psi_k (\sigma) = e^{-ik} \psi_k (\sigma).
\label{tinv}
\ee
It is important to note that $\ctrans^N$ is not the identity operator. 
In fact, 
\be 
\ctrans^N = P = \prod_{i=1}^N \s^z_i,
\ee
the transformed global spin flip operator [\reff{proju}] of Section 
\ref{groundxz}.
The state space may again be decomposed into two subspaces 
corresponding to the eigenvalues $+1$ and $-1$ of $P=\ctrans^N$, 
which we refer to as the even and odd sectors respectively. 
We see that $\ctrans^{2N}=1$, and so the possible values of $k$ are 
$k=\pi j/N$ with $j=0,1,2,\cdots,2N-1$. 
An eigenstate of $\ctrans$ with eigenvalue $e^{-ik}$ will be in the 
even sector if $e^{-ikN}=1$ and in the odd sector if $e^{-ikN}= -1$. 

Almost every quantity depends on $N$, the number of sites. We usually 
suppress this dependence, but in the statement of the following theorem
we make it explicit. 
As we saw in the last section, for even $N$, the lowest eigenvalues in the 
even and odd sectors, which we now denote by $E^N_+$ and $E^N_-$, respectively,
are slightly different.
The expansion of the previous section shows that with our periodic
boundary conditions, they are both equal, up to a correction that is 
exponentially small in $N$, to $N$ times a constant $e_0$, the infinite
volume ground state energy per site.  
We define $E^N_0(k)$ to be $E^N_+$ if $k$ is in the even sector and 
$E^N_-$ if $k$ is in the odd sector. So
\be 
E^N_0(k) = {E^N_+ + E^N_- \over 2} + {E^N_+ - E^N_- \over 2} e^{-ikN}.
\label{ezero_def}
\ee 
For odd $N$ we let $E^N_1(k)$ denote the  
lowest eigenvalue in the subspace of generalized momentum $k$ 
for the Hamiltonian of this section. 
The difference $E^{N+1}_1(k)-E^N_0(k)$ with $N$ even 
is equal to $e_0$ plus the energy of an interface with momentum $k$. 
Our goal is to study this quantity in the 
infinite $N$ limit. If there is a localized interface, then this difference
would be independent of $k$, as explained in the Introduction. 

The quantities $E^{N+1}_1(k)$ and $E^N_0(k)$ are only defined for a
finite set of values of $k$, and the two functions are defined on
different sets of values. To make sense of this difference, we extend
the definitions of these two functions to all $k$. 
The Fourier coefficients $e^N_{0,s}$ are defined by 
\be 
E^N_0(k) = \sum_{s=1}^{2N} e^N_{0,s} \, e^{iks} .
\label{fourier_disp}
\ee 
The RHS of this equation is defined for all $k$, so we can take
it to be the definition of the LHS for all $k$. 
We extend the definition of $E^{N+1}_1(k)$ to all $k$ in the same way.
It is useful to define $e^N_{0,s}$ and $e^{N+1}_{1,s}$
for all $s$ by making them periodic function of 
$s$ with periods $2N$ and $2(N+1)$. Then we can rewrite our Fourier 
series so that they are centered around $s=0$, e.g., 
\be 
E^N_0(k) = \sum_{s=-N+1}^{N} e^N_{0,s} \, e^{iks} .
\label{fourier_dispb}
\ee 
This form is better suited for taking the $N \rightarrow \infty$ limit. 

\begin{theorem}
There exists an $\eps_0 >0$ such that for all $|\eps| < \eps_0$
the following is true:
For $s \in \zed$ there are coefficients $\vareps_s$ such that for all $k$   
\be 
\lim_{N \rightarrow \infty\atop{ N \, {\rm{even}}}}
\left( E^{N+1}_1(k)-E^N_0(k) \right) = \sum_s \vareps_s \, e^{iks} .
\label{disprel}
\ee 
Moreover, there is a constant $c$ such that 
\be
|\vareps_s| \le (c |\epsilon|)^{\lceil |s|/2 \rceil }.
\ee
where the notation $\lceil l \rceil$ denotes the smallest 
integer which is not smaller than $l$. 
We have 
\be
\vareps_2=\vareps_{-2}= \epsilon + O(\epsilon^2).
\label{coeff2}
\ee
So the dispersion relation \reff{disprel} is not a constant function of $k$.
\label{thm_excited}
\end{theorem}

\bigskip
The remainder of this section is devoted to the proof of this theorem.
In the last section we assumed that $N$ was even. It is only for even 
$N$ that the periodic boundary conditions for the original Hamiltonian
\reff{hamorig} lead to the Hamiltonian \reff{eqbu}, and hence to the 
Kirkwood-Thomas equation \reff{eqd}. However, eq. \reff{eqd} is defined
for all $N$ and the proof of the existence of a solution works 
for odd $N$ as well. This allows us to define $E_0^N(k)$ for odd $N$. 
Moreover, the difference between $E_0^N(k)$ and 
$E_0^{N+1}(k)$ converges to a constant $e_0$, the ground state energy per site,
as $N \rightarrow \infty$.
Hence, to prove the theorem we can 
consider the difference $\left(E^N_1(k)-E^N_0(k)\right)$ with $N$ odd. 
Throughout the proof we will
work with this quantity and suppress the superscript $N$. In the
rest of the paper, the $N$--dependence of functions will not be 
explicitly indicated unless needed.

We start by studying what the eigenfunctions of $\ctrans$ look like. 
For $k=\pi j/N$ with $j=0,1,2,\cdots,2N-1$ we define
\be 
\phi_{X,k}(\s) = \sum_{l=1}^{2N} e^{ikl} \, \s_1 \s_2 \cdots \s_l \,
\s(X+l).
\ee
Indices should be taken to be periodic, i.e., 
$\s_{N+i}=\s_i$ for $i=1,2,\cdots,N$.
However, for $l>N$ one should not interpret 
$ \s_1 \s_2 \cdots \s_l $ as $ \s_1 \s_2 \cdots \s_{l-N} $.
Since $\s_i^2=1$, it is  $\s_{l-N+1} \cdots \s_N $.
Note that $\s_1 \s_2 \cdots \s_l \, \s(X+l)=\ctrans^l \s(X)$, so 
we can write the above as 
\be 
\phi_{X,k}(\s) = \sum_{l=1}^{2N} e^{ikl} \, \ctrans^l \, \s(X) ,
\ee
from which it is clear that $\phi_{X,k}(\s)$ is an eigenfunction
of $\ctrans$ with eigenvalue $e^{-ik}$. 

These functions span the subspace of generalized momentum $k$, but they are 
not linearly independent.   
For some choices of $X$ and $k$, $\phi_{X,k}(\s)$ will be zero.
We define the action of $\ctrans$ on a set of sites by 
$\s(\ctrans^l X)=\ctrans^l \s(X)$. More explicitly, we have 
$\ctrans X = \{1\} \d (X+1)$.
Then 
\be 
\phi_{\ctrans^t X,k}(\s)=  \sum_{l=1}^{2N} e^{ikl} \, \ctrans^l \, 
\s(\ctrans^t X) 
=  \sum_{l=1}^{2N} e^{ikl} \, \ctrans^{t+l} \s(X) 
= \sum_{l=1}^{2N} e^{ik(l-t)} \, \ctrans^{l} \s(X) 
= e^{-ikt} \phi_{X,k}(\s).
\label{eqnotind}
\ee
Hence, if two subsets of the lattice are related by a 
generalized translation then the 
corresponding functions are the same up to a multiplicative
constant. If we define two sets $X$ and $Y$ to be equivalent if 
$X=\ctrans^n Y$ for some $n$, then we can partition the subsets of 
$\Lambda$ into equivalence classes. Pick one set from each equivalence
class and let $\xx$ be the resulting collection of subsets of $\Lambda$.
The $\phi_{X,k}$ will still span the subspace of generalized 
momentum $k$ if we only consider $X \in \xx$. 

As we remarked before, the proof of the previous section that the 
Kirkwood--Thomas eq. \reff{eqd} has a solution works for odd $N$ just as 
for even $N$. We let $\Omega(\s)$ be the solution,
\be
\Omega (\s) = \exp \left[ -\frac{1}{2} \sum_Y g(Y) \s (Y)\right].
\ee
This is the ground state of the Hamiltonian in \reff{four0} for odd $N$, 
or equivalently of the Hamiltonian in \reff{four1} with all the $J_j=+1$.
$\Omega(\s)$ is translationally invariant, so if $\psi_k(\s)$ has generalized 
momentum $k$, then $\psi_k(\s)/\Omega(\s)$ does too.
Now suppose that for each $k$ we have an eigenstate $\psi_k(\s)$ 
with momentum $k$.
Then $\psi_k(\s)$  can be written in the form 
\be 
\psi_k(\s) = \Omega(\s) \sum_{X \in \xx} c(X,k) \phi_{X,k}(\s)
= \Omega(\s) 
\sum_{l=1}^{2N} e^{ikl} \, \s_1 \s_2 \cdots \s_l \, \sum_{X \in \xx}
 c(X,k) \, \s(X+l) 
\label{wfin}
\ee
for some coefficients $c(X,k)$, which depend on $k$.  
Let us rewrite the expression for $\psi_k(\s)$ in a manner that 
makes the $k$--dependence more explicit:  
For each $X$ we can write $c(X,k)$ as a Fourier series 
\be 
c(X,k) = \sum_{n=1}^{2N} e^{-ikn} \, e(X,n) .
\label{five8}
\ee
The coefficients $c(X,k)$ are functions of 
$k=\pi j/N$ with $j=0,1,2,\cdots,2N-1$, and hence the sum on
the RHS of \reff{five8} is over $2N$ values (rather than just $N$). 
Using \reff{eqnotind} we have
\be 
\psi_k(\s) = \Omega(\s) \, \sum_{X \in \xx}
\sum_n e(X,n) \, \phi_{\ctrans^n X,k}
 = \Omega(\s) \, \sum_X e(X) \, \phi_{X,k},
\ee
where the coefficients $e(X)$ are defined by the equations 
\be 
e(X)= \sum_{Y \in \xx,n: \ctrans^n Y=X} e(Y,n).
\ee
The wavefunction $\psi_k(\s)$ can now be written in the form
\be 
\psi_k(\s) =  \Omega(\s) 
\sum_{l=1}^{2N} e^{ikl} \, \s_1 \s_2 \cdots \s_l \, \sum_X 
e(X) \, \s(X+l) .
\ee
Note that the $k$--dependence is now entirely contained in the factor
$e^{ikl}$.

We will abbreviate $\langle j,j+1\rangle  \in \boundary X$ by $j:X$ or $X:j$.
Recall that 
$\s_j^x \s_{j+1}^x \s(X) = - \s(X)$ if $j:X$ and it equals $\s(X)$ otherwise.
It easily follows that 
\be
J_j \s_j^x \s_{j+1}^x \s_1 \s_2 \cdots \s_l = s(j,l) 
 \s_1 \s_2 \cdots \s_l, 
\ee
where 
\bea
s(j,l) &=& +1 \quad \hbox{ if} \quad  j\ne l \, {\rm{mod}} \, N\nonumber\\
 &=& -1 \quad \hbox{ if} \quad  j= l \, {\rm{mod}} \, N.
\label{sjn}
\eea
Thus
\begin{eqnarray}
(H \psi_k)(\s) 
 &=& \Omega (\s) \sum_{l=1}^{2N} e^{ikl} 
\s_1 \s_2 \cdots \s_l \,\nonumber\\ 
&&\, \Bigl[-\sum_{j=1}^N \exp[\sum_{Y:j} g(Y) \s (Y)] \, s(j,l)
\, \Bigl( \sum_X e(X) \s (X+l) -
2 \sum_{X :j-l} e(X) \s (X+l)\Bigr) \nonumber\\ 
&& + \epsilon \sum_{j=1}^N \s_j \s_{j+1} \sum_X e(X) \s (X+l) \Bigr].
\label{lhs}
\end{eqnarray}
The above must equal $E_1(k) \psi_k(\s)$. 
Canceling the common factor of $\Omega(\s)$, the Schr\"odinger
equation for the Hamiltonian $H$ [\reff{four1}] becomes 
\bea
&&
\sum_{l=1}^{2N} e^{ikl} 
\s_1 \s_2 \cdots \s_l \, \Bigl[-\sum_{j=1}^N \exp[\sum_{Y:j} g(Y) \s (Y)]
\, s(j,l) 
\,  \Bigl( \sum_X e(X) \s (X+l) -
2 \sum_{X :j-l} e(X) \s (X+l)\Bigr) \nonumber\\
&& + \epsilon \sum_{j=1}^N \s_j \s_{j+1} \sum_X e(X) \s (X+l) 
- E_1 (k)\sum_X e(X) \s (X+l) \Bigr] =0
\label{intf}
\eea
If eq.\reff{intf} was of the form  
\be
\sum_{l=1}^{2N} e^{ikl} f(l,\s) =0 \label{fourier}
\ee
then we would have been able to conclude that $f(l,\s)=0$ for all $l$. 
However, even though eq.\reff{intf} resembles \reff{fourier}, the
two equations are not quite identical in form. This is
 because $E_1(k)$ depends on $k$. 
To cast \reff{intf} in the form \reff{fourier}, 
we write $E_1(k)$ as a Fourier series in $k$.  
When $\epsilon=0$, $E_1(k) - E_0(k) = 2$. So we write it as 
\be 
E_1(k) = E_0(k) + 2 + \sum_{s=1}^{2N} e_s e^{-iks}
\label{ek}
\ee
Using the definition of $E_0(k)$ [eq. \reff{ezero_def}],
\bea
\sum_{l=1}^{2N} e^{ikl} 
\s_1 \cdots \s_l \, E_1 (k)
\sum_X e(X) \s (X+l) 
&=& \left(2 + {E_+ + E_- \over 2}\right) \sum_{l=1}^{2N} e^{ikl} 
\s_1 \cdots \s_l \sum_X e(X) \s(X+l) \nonumber\\ 
&&\,\, + \sum_{l,s=1}^{2N} e^{ikl} \s_1 \cdots \s_{l+s} \, e_s
\sum_X e(X) \s(X+s+l)\nonumber\\
&& \,\, + {E_+ - E_- \over 2 } e^{-ikN} \sum_{l=1}^{2N} e^{ikl} 
\s_1 \cdots \s_l \sum_X e(X) \s(X+l) 
\nonumber\\ \label{energyfs}
\eea
where we have made a change of variables 
$l \ra l+s$.
In the expression $\s_1 \cdots \s_{l+s}$ the index $l+s$ can be 
as large as $4N$. For $i=1,2,\cdots,N$, we interpret $\s_{i+N}$, 
$\s_{i+2N}$ and $\s_{i+3N}$ to all be $\s_i$. 
By making a change of variables $l \ra l+N$, and using 
$\s_{l+1} \cdots \s_{l+N} = \s(\Lambda)$ and $\s(X+l+N)=\s(X+l)$, 
we rewrite the last term on the RHS of \reff{energyfs} as 
follows:
\be 
{E_+ - E_- \over 2 } e^{-ikN} \sum_{l=1}^{2N} e^{ikl} 
\s_1 \cdots \s_l \sum_X e(X) \s(X+l) 
= {E_+ - E_- \over 2 } \s(\Lambda) \sum_{l=1}^{2N} e^{ikl} 
\s_1 \cdots \s_l \sum_X e(X) \s(X+l).
\ee

If we use \reff{energyfs} in \reff{intf} the resulting equation is 
of the form \reff{fourier}. Hence, after canceling a common factor 
of $\s_1 \s_2 \cdots \s_l$, we conclude that
\bea
&&
-\sum_{j=1}^N \exp[\sum_{Y:j} g(Y) \s (Y)]
\, s(j,l) 
\,  \Bigl( \sum_X e(X) \s (X+l) -
2 \sum_{X :j-l} e(X) \s (X+l)\Bigr) \nonumber\\
&& + \epsilon \sum_{j=1}^N \s_j \s_{j+1} \sum_X e(X) \s (X+l) 
- (2+{E_+ + E_- \over 2}) \sum_X e(X) \s (X+l) 
 \nonumber\\
&&
- \sum_{s=1}^{2N} \s_{l+1} \cdots \s_{l+s} \, e_s \sum_X e(X) \s(X+s+l) 
- {E_+ - E_- \over 2} \s(\Lambda) \sum_X e(X) \s (X+l) 
= 0. \nonumber\\ 
\label{intfb}
\eea
Recall that the coefficients $g(Y)$ satisfy eq.\reff{eqdd}:
\be
  - \sum_{j=1}^N \exp(\sum_{Y : j} g(Y) \s(Y))  
 + \epsilon \sum_{j=1}^N \, \s_j \s_{j+1} = {E_+ + E_- \over 2} 
 + {E_+ - E_- \over 2} \s(\Lambda).
\label{fpe1} 
\ee
Multiplying this equation by $\sum_X e(X) \s(X+l)$ and 
subtracting the result from 
\reff{intfb}
\bea
&&
\sum_{j=1}^N \exp[\sum_{Y:j} g(Y) \s (Y)]
\,  \Bigl[ (1-s(j,l)) \, \sum_X e(X) \s (X+l) 
+ 2 s(j,l) \, \sum_{X :j-l} e(X) \s (X+l) \Bigr] 
\nonumber\\
&& 
- 2 \sum_X e(X) \s (X+l) 
- \sum_{s=1}^{2N} \s_{l+1} \cdots \s_{l+s} \, e_s 
\sum_X e(X) \s(X+s+l) 
= 0. \nonumber\\ 
\label{eqdif}
\eea
Defining $h(Y)$ by 
\be 
\exp(\sum_{Y : N} g(Y) \s(Y))  = 1+ \sum_Y h(Y) \s(Y)
\label{eqdefh}
\ee
we have
\be 
h(Y) = \sum_{n=1}^\infty {1 \over n !} 
\sum_{Y_1, \cdots, Y_n : N, \Delta=Y} 
g(Y_1) \cdots g(Y_n).
\label{hdef}
\ee
Using the translation invariance of the $g(Y)$
\bea 
\exp(\sum_{Y : j} g(Y) \s(Y))  &=& 
\exp(\sum_{Y : N} g(Y+j) \s(Y+j))  \nonumber \\
&=& \exp(\sum_{Y : N} g(Y) \s(Y+j))  = 
 1+ \sum_Y h(Y) \s(Y+j).
\label{hy}
\eea
Inserting \reff{hy} in \reff{eqdif} we have 
\bea
&&
\sum_{j=1}^N  
\,  \Bigl[ (1-s(j,l)) \, \sum_X e(X) \s (X+l) 
+ 2 s(j,l) \, \sum_{X :j-l} e(X) \s (X+l) \Bigr] 
\nonumber\\
&& 
+ \sum_{j=1}^N  \sum_Y h(Y) \s(Y+j)
\,  \Bigl[ (1-s(j,l)) \, \sum_X e(X) \s (X+l) 
+ 2 s(j,l) \, \sum_{X :j-l} e(X) \s (X+l) \Bigr] 
\nonumber\\
&& 
- 2 \sum_X e(X) \s (X+l) 
- \sum_{s=1}^{2N} \s_{l+1} \cdots \s_{l+s} \, e_s 
\sum_X e(X) \s(X+s+l) 
= 0. \nonumber\\ 
\label{withl}
\eea

Eq. \reff{withl} must hold for all $l$ and $\sigma$. The equations for 
different values of $l$ are in fact identical. To see this  
we make a change of variables $j \rightarrow j+l$ in the sums over $j$.
Note that $s(j+l,l)=s(j,N)$. The resulting 
equation must hold for all configurations $\sigma$. Hence, 
we can also replace $\sigma$ by the configuration 
obtained by translating $\sigma$ by $l$ sites so 
that $\sigma(X+l)$ becomes $\sigma(X)$. The result of these two 
changes of variables is that, for each value of $l$, eq. \reff{withl}
reduces to the following equation, which is the $l=N$ case of 
eq. \reff{withl}:

\bea
&&
\sum_{j=1}^N  
\,  \Bigl[ (1-s(j,N)) \, \sum_X e(X) \s (X) 
+ 2 s(j,N) \, \sum_{X :j} e(X) \s (X) \Bigr] 
\nonumber\\
&& 
+ \sum_{j=1}^N  \sum_Y h(Y) \s(Y+j)
\,  \Bigl[ (1-s(j,N)) \, \sum_X e(X) \s (X) 
+ 2 s(j,N) \, \sum_{X :j} e(X) \s (X) \Bigr] 
\nonumber\\
&& 
- 2 \sum_X e(X) \s (X) 
- \sum_{s=1}^{2N} \s_{1} \cdots \s_{s} \, e_s 
\sum_X e(X) \s(X+s) 
= 0. \nonumber\\ 
\label{s3}
\eea
Note that 
\bea
\sum_j s(j,N) \sum_{X:j} e(X) \s (X) &=&
\sum_X e(X) \s (X) \sum_{j:X} s(j,N)\nonumber\\
&=& \sum_X n(X)  e(X) \s (X),
\label{f19}
\eea
where we have defined
\be
n(X) := \sum_{j:X} s(j,N),
\label{nx}
\ee
and the sum is over $j$ such that $\langle j,j+1\rangle  \in \boundary X$. 
Note that $n(X)$ is either zero or an even integer.
Moreover,
\bea
1-s(j,N) &=& 2 \quad {\hbox{if}} \quad j=N\nonumber\\
&=& 0 \quad {\hbox{if}} \quad j\ne N.
\label{prop}
\eea
Hence, eq. \reff{s3} can be written as
\bea
&&
2 \, \sum_X n(X) \, e(X) \s (X) 
+ 2 \sum_Y h(Y) \s(Y) \, \sum_X e(X) \s (X) 
\nonumber\\
&& 
+ 2 \sum_Y \, \sum_X \, \sum_{j: X}  \, h(Y) \s(Y+j) \, 
s(j,N) \, e(X) \s (X)
\nonumber\\
&& 
- \sum_{s=1}^{2N} \s_{1} \cdots \s_{s} \, e_s 
\sum_X e(X) \s(X+s) 
= 0. \nonumber\\ 
\label{s7}
\eea
Recall that $j:X$ means that exactly one of the sites 
$j$ and $j+1$ is in $X$.
Define $j::X$ as follows: If $j \ne N$, $j::X$ means the same as 
$j:X$. However, $N::X$ means either both of the sites $N$ and $1$ 
are in $X$ or both are not. This is a natural definition since the 
sites $j$ for which $j::X$ are precisely the sites for which there is 
an interface between the sites $j$ and $j+1$. 
With this definition, 

\bea
2 \sum_Y h(Y) \s(Y) \, \sum_X e(X) \s (X) 
&+& 2 \sum_Y \, \sum_X \, \sum_{j: X}  \, h(Y) \s(Y+j) \, 
s(j,N) \, e(X) \s (X) \nonumber\\ 
&=&  2 \sum_Y \, \sum_X \, \sum_{j:: X}  \, h(Y) \s(Y+j) \, e(X) \s (X).
\eea
Since 
\be
\s_1 \cdots \s_s \, \s(X+s) = \s(\ctrans^s X),
\label{tsx}
\ee
the last term in \reff{s7} can be written  as 
\be 
\sum_{s=1}^{2N} e_s \sum_X e(X) \s(\ctrans^s X) 
=\sum_{s=1}^{2N} e_s \sum_X e(\ctrans^{-s} X) \s(X), 
\label{eqtrans}
\ee
where the equality follows by a change of variables in the sum.  
(Since $\ctrans^{2N}=1$, $\ctrans^{-s}= \ctrans^{2N-s}$.)
Thus \reff{s7} holds for all configurations $\s$ if and only if for all $X$, 
\be
2  n(X)  e(X)  +  2 \sum_{Y,Z,j:: Z} h(Y) e(Z) 1((Y+j) \d Z=X)
=  \sum_{s=1}^{2N} e_s \, e(\ctrans^{-s} X).
\label{eqfpe}
\ee

The integer $n(X)$ is zero for sets of the form $X=\{1,2,\cdots,s\}$
and $X=\{s+1,s+2,\cdots,N\}$. These are the sets $\ctrans^m \emptyset$ 
where $m=0,1,\cdots,2N-1$. 
Let us assume that $e(X)=0$ for all $X$ for which $n(X)=0$, except
for $X=\emptyset $ for which it is equal to unity.
This is essentially a normalization condition.
(A priori there is no reason that a solution with these properties must
exist, but we will show that it does.) 
With this assumption, if $X=\ctrans^m \emptyset$ then 
\be
\sum_{s=1}^{2N} e_s \, e(\ctrans^{-s} X) = e_m.
\ee
Thus eq. \reff{eqfpe} gives
\bea
 e_m= 2 \sum_{Y,Z,j:: Z} h(Y) e(Z) 1((Y+j) \d Z=\ctrans^m \emptyset).
\label{es2}
\eea

For $X$ for which $n(X) \ne 0$, we obtain the relation  
\be
e(X) = { 1 \over 2n(X)} \biggl[  
 -  2 \sum_{Y,Z,j:: Z} h(Y) e(Z) 1((Y+j) \d Z=X)
 + \sum_{s=1}^{2N} e_s \, e(\ctrans^{-s} X) \biggl].
\label{ex}
\ee
We will show that these equations [\reff{es2} and \reff{ex}]
have a solution by writing them as a 
fixed point equation. 
Consider the set of variables 
\be e := \{e(X): n(X) \ne 0\} \cup \{e_s: s=1,2,\cdots,2N \}.
\ee
Equations \reff{es2} and \reff{ex} form a fixed point equation 
for $e$
\be
\fp(e)=e.
\label{fpeaf}
\ee

Let us introduce the norm 
\be
||e|| := \sum_{l=1}^{2N} |e_l| \weight^{-w_N(\ctrans^l \emptyset)} 
+ 2 \sum_{X \atop{n(X) \ne 0}} |e(X)| n(X) \weight^{-w_N(X)},
\label{norm}
\ee
where $w_N(X)$ is the number of  bonds in the smallest set of  bonds
which contains $X$ and intersects the bond $\langle N,1\rangle $ 
and which is connected in the 
sense used in the previous section to define $w(X)$ [see the discussion
after \reff{normg}]. 
The factor of 2 in the norm is included merely for later convenience.
\smallskip

We prove that the fixed point equation for $e$ has a solution 
by using the contraction mapping theorem as we did in the previous 
section. We must show that there is a $\delta'>0$ such that 
\be
||\fp(e)-\fp({\widetilde{e}})|| \le \frac{1}{2}||e - {\widetilde{e}}|| \quad {\hbox{for}} \quad
||e||, ||{\widetilde{e}}|| \le \delta' ;
\label{exboundone}
\ee
\be
||\fp(e)|| \le \delta' \quad {\hbox{for}} \quad
||e|| \le \delta'. 
\label{exboundtwo}
\ee
To verify \reff{exboundone}, we use  
\reff{es2} and \reff{ex} to see that 
\bea
 ||\fp(e) - \fp({\widetilde{e}})|| 
&\le & 2  \sum_Y \sum_Z \sum_{j::Z}
|h(Y)|\,|e(Z) -{\widetilde{e}}(Z)|  \weight^{-w_N((Y+j) \d Z)}
\nonumber\\
&& + \sum_s \sum_{X : n(X) \ne 0 } \,
|e_s \, e(\ctrans^{-s}X) - {\widetilde{e}}_s \, {\widetilde{e}}(\ctrans^{-s}X)| \weight^{-w_N(X)}.
\label{bound}
\eea
To continue we need the following two inequalities.
\be
w_N((Y+j) \d Z) \le w_N(Y)+w_N(Z), \quad {\rm for} \quad j::Z
\label{wtria}
\ee
\be
w_N(X) \le w_N(\ctrans^{-s} X) + w_N(\ctrans^s \emptyset) 
\label{wtrib}
\ee
The inequality \reff{wtrib} can equivalently be written as
\be
w_N(\ctrans^s X) \le w_N(X) + w_N(\ctrans^s \emptyset) .
\label{wtrib2}
\ee

In the following proofs of these inequalities, ``a connected set of 
 bonds'' will always mean connected in the sense used to define 
$w_N(X)$.
To prove \reff{wtria}, let $A$ and $B$ be connected sets of 
 bonds which contain $Y$ and $Z$ respectively, both of which intersect
the bond $\langle N,1\rangle $, and such that $w_N(Y)=|A|$, $w_N(Z)=|B|$.
We consider the cases of $j=N$ and $j \ne N$ separately. 
First let $j=N$. 
Then $A \cup B$ is a connected set of  bonds which contains 
$(Y+j) \d Z = Y \d Z$ and intersects the bond $\langle N,1\rangle $. So 
\be
w_N((Y+j) \d Z) \le |A \cup B| \le |A| + |B| = w_N(Y)+w_N(Z).
\ee
Now suppose $j \ne N$. Then $j::Z$ means that either $j$ or $j+1$ is 
in $Z$ and so is in $B$. Since $\langle N,1\rangle $ intersects $A$, 
the set $A+j$ contains at least one of the sites $j$ and $j+1$.
Thus $(A+j) \cup B$ is a connected set of  bonds. It contains 
$(Y+j) \d Z$ and intersects the bond $\langle N,1\rangle $. So 
\be
w_N((Y+j) \d Z) \le |(A+j) \cup B| \le |A+j| + |B| = w_N(Y)+w_N(Z).
\ee
This proves \reff{wtria}.
The inequality \reff{wtrib2} is a special case of \reff{wtria}.
To see this, note that 
\be
\ctrans^s X = (X+s)\, \d \,\ctrans^s \emptyset,
\ee
so if we take $Y=X$, $Z= \ctrans^s \emptyset$ and $j=s$, then 
\reff{wtria} becomes \reff{wtrib2}.
(It is easy to check that $s::\ctrans^s \emptyset$ for all $s$.)

We will also need the relation,
\be
|\{j: \, j::Z \}| = n(Z)+1 .
\label{nz}
\ee
Recalling the definition of $n(Z)$ [\reff{nx}], and of $s(j,N)$ 
[\reff{sjn}], 
\be
1+n(Z) = 1+ \sum_{j:Z} s(j,N) = 1 - {\mathbf{1}}(N:Z) + 
\sum_{j:Z, j  \ne N} 1,
\ee
where ${\mathbf{1}}(\cdot)$ denotes an indicator function.
Now $j:Z$ and $j::Z$ are equivalent if $j \ne N$. 
Moreover, $N::Z$ holds if and only if $N:Z$ does not hold.
So $\left(1 - {\mathbf{1}}(N:Z)\right)={\mathbf{1}}(N::Z)$. 
This proves \reff{nz}.

Using \reff{wtria} and \reff{nz}, the first term in \reff{bound} is 
\bea
&\le& 2  \sum_Y \sum_Z \sum_{j::Z}
|h(Y)| \weight^{-w_N(Y)} \,|e(Z) -{\widetilde{e}}(Z)| \weight^{-w_N(Z)} 
\nonumber\\
& = & 2  \sum_Y \sum_Z 
[n(Z)+1] |h(Y)| \weight^{-w_N(Y)} \,|e(Z) -{\widetilde{e}}(Z)| \weight^{-w_N(Z)} 
\label{bound2}
\eea
If $n(Z)=0$ then either both of $e(Z)$ and ${\widetilde{e}}(Z)$ are $0$, or both are $1$. 
So $|e(Z) -{\widetilde{e}}(Z)|=0$ when $n(Z)=0$. 
Thus we can bound $\left(n(Z)+1\right)$ by $2 n(Z)$ 
on the RHS of \reff{bound2}. Hence,
\be
{\hbox{RHS of }} \reff{bound2}  \le  2  \sum_Y  |h(Y)| \weight^{-w_N(Y)} \,||e -{\widetilde{e}}||.  
\label{bounda}
\ee

Using \reff{wtrib} and the triangle inequality in the form
\be
|e_s \, e(\ctrans^{-s}X) - {\widetilde{e}}_s \, {\widetilde{e}}(\ctrans^{-s}X)| \le 
 |e_s| \, |e(\ctrans^{-s}X)-{\widetilde{e}}(\ctrans^{-s}X)|
+  |e_s-{\widetilde{e}}_s| \, |{\widetilde{e}}(\ctrans^{-s}X)|,
\label{bound3}
\ee
the second term in \reff{bound} is bounded by 
\bea
&& \sum_s \sum_{X : n(X) \ne 0 } \, |e_s| \weight^{-w_N(\ctrans^s \emptyset)}
\, |e(\ctrans^{-s}X)-{\widetilde{e}}(\ctrans^{-s}X)| \weight^{-w_N(\ctrans^{-s} X)}
\nonumber\\
&+&  \sum_s \sum_{X : n(X) \ne 0 } \, |e_s-{\widetilde{e}}_s| 
\weight^{-w_N(\ctrans^s \emptyset)}
\, |{\widetilde{e}}(\ctrans^{-s}X)| \weight^{-w_N(\ctrans^{-s} X)}.
\nonumber\\
&\le& {1 \over 2} (||e|| \, ||e-{\widetilde{e}}|| + ||e-{\widetilde{e}}|| ||{\widetilde{e}}|| )
\le  \delta' ||e-{\widetilde{e}}||, 
\label{boundb}
\eea
since $||e||$ and $||{\widetilde{e}}||$ are no greater than $\delta'$.

Using the above inequalities \reff{bounda} and \reff{boundb}, we have
\be
 ||\fp(e) - \fp({\widetilde{e}})|| 
\le  [2  \sum_Y  |h(Y)| \weight^{-w_N(Y)} +\delta'] \,||e -{\widetilde{e}}|| .
\ee
It is easily shown that 
\be
w_N(X_1 \d \cdots \d X_n) \le \sum_{k=1}^n w_N(X_k) .
\label{wtrina}
\ee
So using \reff{hdef}
\be 
\sum_Y |h(Y)| \weight^{-w_N(Y)} \le \sum_{n=1}^\infty {1 \over n !} 
\sum_{Y_1 : N,\ldots,Y_n : N}
\prod_{k=1}^n \weight^{-w_N(Y_k)}|g(Y_k)|.
\label{wtrinb}
\ee
The constraint $Y_k:N$ implies that $Y_k$ intersects the bond $\langle N,1\rangle $, 
and so $w_N(Y_k)=w(Y_k)$. Hence, 
\be 
{\hbox{RHS of }} \reff{wtrinb} = \sum_{n=1}^\infty {1 \over n !} 
\sum_{Y_1 : N,\ldots,Y_n : N}
\prod_{k=1}^n \weight^{-w(Y_k)}|g(Y_k)|
=e^{||g||}-1 \le e^\delta-1.
\label{wtrinbcont}
\ee
The last inequality follows from Theorem \ref{theo1}.
So 
\be
||\fp(e) - \fp({\widetilde{e}})|| \le  K \, ||e - {\widetilde{e}}||.
\ee
where 
\be
K = 2 (e^\delta -1) + \delta'.
\ee
If $\delta$ and $\delta'$ are small enough, then $K \le 1/2$. 

To prove \reff{exboundtwo}, we use \reff{es2} and 
\reff{ex} to compute
$\fp(0)$. Note that $e=0$ means that $e_s=0$ for all $1\le s \le 2N$,
and $e(X)=0$ for all $X$ except
$X=\emptyset $. We always have $e(\emptyset )=1$.
Letting ${\widetilde{e}}$ denote $F(0)$, we have 
\be
{\widetilde{e}}_m = 2 h(\ctrans^m \emptyset), 
\label{abv}
\ee
and for $X$ with $n(X) \ne 0$ 
\be 
{\widetilde{e}}(X) =  { 1 \over 2n(X)} 
 \, [- 2 h(X)].
\ee
Thus
\be
||\fp(0)|| \le 2 \sum_Y  |h(Y)| \weight^{-w_N(Y)} 
\le 2(e^\delta-1).
\ee
If we decrease $\delta$, then $K$ decreases. Hence, we can assume 
$\delta$ to be small enough so that
$ 2(e^\delta-1) < \delta'/2 $. So 
\be
||\fp(e)|| \le ||\fp(e)-\fp(0)|| + ||\fp(0)||
\le {1 \over 2} ||e|| + 2(e^\delta-1) \le \delta'
\ee
since $||e|| \le \delta'$.

This finishes the proof that the fixed point equation has 
a solution and thus completes the construction of eigenstates 
of $H$ with generalized momentum $k$. 
When $\epsilon=0$ these states are the lowest eigenstates in the 
subspaces of generalized momentum $k$ for $k \ne 0$, and the next to 
lowest for $k=0$. 
The same sort of argument that was used in Section \ref{groundxz}
proves that this is true for all $\epsilon$ such that $|\epsilon|<\epsilon_0$,
for some $\epsilon_0 >0$. We refer the reader to section 3 of \cite{DK} for a completely 
analogous argument. 

We now consider the convergence of the $N \ra \infty$ limit.
We start by asking how the volume $\Lambda$ enters the ground state 
fixed point equation \reff{fixedpt}. The sets $X_i$ in this equation are
subsets of $\Lambda$ and the definition of nearest neighbor for the 
term $1_{\nn}(X)$ depends on $\Lambda$. The solution $g$ of eq. \reff{fixedpt}
will depend on $\Lambda$, and so we denote it by $g_\Lambda$.
However, we can consider this equation for the infinite lattice $\zed^d$. 
This means that the sets can be any finite subset of $\zed^d$, and nearest
neighbor is defined in the usual  way for $\zed^d$. The proof of the 
ground state section shows that this infinite volume fixed point 
equation has a solution, which we denote by $g_\infty$. 
One can prove that $g_\Lambda$ converges to $g_\infty$ in an appropriate 
sense by showing $g_\Lambda$ is an approximate solution of the fixed 
point equation that defines $g_\infty$. We refer the reader to 
\cite{DK} for details. 

The fixed point equations, \reff{es2} and \reff{ex}, of this section 
can also be defined for the infinite lattice $\zed^d$, and the fixed point
argument of this section proves it has a solution. This solution 
includes the Fourier coefficients $\vareps_s$, so in this way the 
coefficients $\vareps_s$ of the Theorem \ref{thm_excited} are defined. 
The convergence of 
$E^{N+1}_1(k)-E^N_0(k)$ to $\sum_s \vareps_s \, e^{iks}$ can be proved by 
the methods of \cite{DK} as well.

The last step in the proof is to show that $e_2$ and $e_{-2}=e_{2N-2}$
are not zero in the infinite length limit. 
We start with \reff{fixedpt} to compute $g$ to first order in $\epsilon$. 
At first order in $\epsilon$ the only nonzero coefficients $g(X)$ are
for sets $X$ which consist of a pair of adjacent sites. In this 
case $g(X)=\epsilon/2 + O(\epsilon^2)$.
By \reff{hdef}, the only $Y$ for which $h(Y)$ is nonzero at first order
in $\epsilon$ is a set of nearest neighbor sites satisfying $N:Y$.  
There are two such sets, $\{1,2\}$ and $\{N-1,N\}$.  
They have $h(Y)=\epsilon/2 + O(\epsilon^2)$.
Now consider eq.\reff{es2}. $h(Y)$ is always at least first order in 
$\epsilon$, but there is one $Z$ for which $e(Z)$ is zeroth order in 
$\epsilon$, namely, $e(\emptyset)=1$.
For this $Z$ the only $j$ satisfying $j::Z$ is $j=N$.  
Thus the first order contribution to $e_m$ is of the form
\be
2 \sum_{Y} h(Y) 1(Y =\ctrans^m \emptyset).
\ee
The sets $\{1,2\}$ and $\{N-1,N\}$ are of the form 
$\ctrans^m \emptyset$ for $m=2$ and $m=2N-2$, respectively. 
Thus 
\be
e_2=e_{2N-2}= \epsilon + O(\epsilon^2).
\ee
This proves \reff{coeff2} of Theorem \ref{thm_excited}.

\section{Antiferromagnetic XXZ Chain}
\label{af}

In this section we study the antiferromagnetic XXZ model whose Hamiltonian
on the 1-dimensional lattice $\Lambda = \{1,2,\ldots N\}$ is 
\be
{\widetilde{H}} = \sum_{j=1}^N \s^z_j \s^z_{j+1} 
+ \epsilon \sum_{j=1}^N (\s^x_j \s^x_{j+1} + \s^y_j \s^y_{j+1}) 
\ee
Using $\sigma^y = i \sigma^x \sigma^z$ we have
\be
{\widetilde{H}} = \sum_{j=1}^N \s^z_j \s^z_{j+1} 
+ \epsilon \sum_{j=1}^N \s^x_j \s^x_{j+1}(1 - \s^z_j \s^z_{j+1}) 
\label{oraf}
\ee
As before consider a unitary operator that causes
a rotation about the $Y$--axis in spin space:
$\displaystyle{\fun:= \exp \left(i\frac{\pi}{4}\sum_{j \in \Lambda} 
\sigma^y_j\right)}$ 
so that 
\be
\fun {\widetilde{H}} \fun^{-1} = \sum_{j=1}^N \s^x_j \s^x_{j+1} 
+ \epsilon \sum_{j=1}^N \s^z_j \s^z_{j+1}(1 - \s^x_j \s^x_{j+1}) 
\ee
For the antiferromagnet we proceed as in the previous section and 
use the unitary transformation $\sun$ [eq. \reff{sundef}]: 
\be
H= \sun \fun {\widetilde{H}} \fun^{-1} \sun^{-1}
 = - \sum_{j=1}^N J_j \s^x_j \s^x_{j+1} 
+ \epsilon \sum_{j=1}^N \s^z_j \s^z_{j+1}(1 + J_j \s^x_j \s^x_{j+1}) 
\label{hxxz}
\ee
where $J_j=1$ for $j \ne N$ and $J_N$ is $1$ when $N$ is even and 
$-1$ when $N$ is odd. 
$\widetilde{H}$ is translation invariant and commutes with the 
global spin flip operator ${\widetilde{P}}$ [\reff{ptilde}]. 
So $H$ commutes with $\ctrans$ [\reff{tee}], as it did in the previous
section.

When $N$ is even (and so $J_j=+1 \, \forall\, j$ ), the ground state wave function 
\be
\Omega (\s) = \exp \left[ -\frac{1}{2} \sum_Y g(Y) \s (Y)\right],
\ee
must satisfy
\be
 - \sum_j \exp\Bigl[\sum_{X:j} g(X) \s(X)\Bigr] 
+ \epsilon \sum_j \s_j \s_{j+1} \left[1 + \exp\Bigl[\sum_{X:j} g(X) \s(X)\Bigr]
 \right] = {E_+ + E_- \over 2} + {E_+ - E_- \over 2} \sigma(\Lambda)
\label{xxzgs}
\ee
where $X:j$ means $\langle j,j+1\rangle  \in X$.
Theorem \ref{theo1} of Section \ref{groundxz} holds 
for this model. We omit the proof since it is analogous 
to the proof in Section \ref{groundxz}. Since the dimension $d=1$, we 
choose $\delta = 4M^{-1}$ as given by \reff{del}. 

To study interfaces in this model we take $N$ to be odd. So $J_N=-1$. 
We recall that in this case, $\ctrans = \s^z_1 \trans$ [\reff{genT}].
As before we look for a solution of the form 
\be 
\psi_k(\s) =  \Omega(\s) 
\sum_{l=1}^{2N} e^{ikl} \, \s_1 \s_2 \cdots \s_l \, \sum_X 
e(X) \, \s(X+l) .
\ee
The Schr\"odinger equation $(H \psi_k)(\s) = E_1(k) \psi_k(\s)$, becomes
(after canceling a common factor of $\Omega(\s)$)
\bea
&&
\sum_{l=1}^{2N} e^{ikl} 
\s_1 \s_2 \cdots \s_l \, \Bigl[-\sum_{j=1}^N \exp[\sum_{Y:j} g(Y) \s (Y)]
\, s(j,l) 
\,  \Bigl( \sum_X e(X) \s (X+l) 
- 2 \sum_{X :j-l} e(X) \s (X+l)\Bigr) 
\nonumber\\
&& + \epsilon \sum_{j=1}^N \s_j \s_{j+1} \sum_X e(X) \s (X+l) 
\nonumber\\
&& + \epsilon \sum_{j=1}^N \s_j \s_{j+1} \exp[\sum_{Y:j} g(Y) \s (Y)]
\, s(j,l) \, \Bigl( \sum_X e(X) \s (X+l) 
- 2 \sum_{X :j-l} e(X) \s (X+l)\Bigr) 
\nonumber\\
&& - E_1 (k)\sum_X e(X) \s (X+l) \Bigr] =0.
\label{intfxxz}
\eea
This is the analog of \reff{intf}. 
We now proceed by analogy with the derivation of 
\reff{es2} and \reff{ex} from  \reff{intf}. 
This leads to the equation:
\bea
&&
 2 \sum_X n(X) \, e(X) \s (X)
- 2 \epsilon \s_N \s_1 \, \sum_X e(X) \s (X) 
 \nonumber\\
&& 
- 2 \epsilon \sum_{j=1}^N \s_j \s_{j+1} \, s(j,N) \, 
\sum_{X :j} e(X) \s (X)
- \sum_{s=1}^{2N} \s_1 \cdots \s_s \, e_s 
\sum_X e(X) \s(X+s) 
\nonumber\\
&&  + 2 \sum_Y h(Y) \, \s(Y) \sum_X e(X) \, \s(X)
+2\,\sum_{j=1}^N \sum_Y h(Y) \, \s(Y+j) \, s(j,N) 
  \, \sum_{X :j} e(X) \s (X)
\nonumber\\
&& + \epsilon \sum_{j=1}^N \s_j \s_{j+1} \sum_Y h(Y) \s(Y+j)
\, (s(j,N)-1) \, \sum_X e(X) \s (X) 
 \nonumber\\
&& 
- 2 \epsilon \sum_{j=1}^N \s_j \s_{j+1} \sum_Y h(Y) \, \s(Y+j)
 \, s(j,N) \, \sum_{X :j} e(X) \s (X)
 \nonumber\\
&&
= 0 \nonumber\\ 
\label{intffferro}
\eea

Eq. \reff{intffferro} yields the 
following equations which are the analogs of \reff{es2} and \reff{ex}.
For $X$ for which $n(X) \ne 0$ we have
\bea
e(X) &=& { 1 \over 2n(X)} \biggl[ 
2\epsilon \sum_{j::Z} \,\, \sum_Z e(Z) 1(Z \d \{j,j+1\} = X) 
\nonumber\\
& &  + \sum_{s=1}^{2N} e_s \, e(\ctrans^{-s} X) 
- 2 \sum_{j::Z} \,\, \sum_{Y,Z}   h(Y) e(Z) 1((Y+j) \d Z=X)
\nonumber\\
& &  + 2\epsilon \sum_{j::Z} \,\, \sum_{Y,Z}   
h(Y) e(Z) 1(Z\d(Y+j) \d \{j,j+1\}=X) \biggr].
\nonumber\\
\label{exaf}
\eea
Recall that $j::Z$ means that exactly one of $j$ and $j+1$ is in $Z$ 
if $j \ne Z$, and $N::Z$ means that either both of $N$ and $1$ are in $Z$ 
or neither of them is.
For $X=\ctrans^m \emptyset$ we have
\bea
e_m &=& 
2\epsilon \sum_{j::Z} \,\, \sum_Z e(Z) 1(Z \d \{j,j+1\}= \ctrans^m \emptyset) 
\nonumber\\
& & + 2\epsilon \sum_{j::Z} \,\, \sum_{Y,Z}   
h(Y) e(Z) 1(Z\d(Y+j) \d \{j,j+1\}= \ctrans^m \emptyset) 
\nonumber\\
& & - 2 \sum_{j::Z} \,\, \sum_{Y,Z}   
h(Y) e(Z) 1((Y+j) \d Z= \ctrans^m \emptyset).
\nonumber\\
\label{emaf}
\eea
Recall that $n(X)=0$ if and only if $X$ is of the form $\ctrans^m(\emptyset)$
for some integer $m$. As in the previous section, 
we assume $e(\emptyset)=1$ and 
$e(\ctrans^m(\emptyset))=0$ for $m\ne 0$. 

We let $e$ denote the same collection of variables as in the previous
section and continue to use the norm \reff{norm}.
Equations \reff{exaf} and \reff{emaf} form a fixed point equation
which can be written as $ \fp(e)=e$.
(Of course, the function $\fp$ is different from the $\fp$ of the 
previous section.) We prove there is a solution to the fixed point
equation by proving \reff{exboundone} and \reff{exboundtwo}.

To prove \reff{exboundone} we use \reff{exaf} and \reff{emaf} to see 
that 
\bea
 ||\fp(e) - \fp({\widetilde{e}})|| 
&\le & 2 |\epsilon|  \sum_Z \sum_{j::Z}
|e(Z) -{\widetilde{e}}(Z)|  \weight^{-w_N(Z \d \{j,j+1\})}
\nonumber\\
&& + \sum_s \sum_{X : n(X) \ne 0 } \,
|e_s \, e(\ctrans^{-s}X) - {\widetilde{e}}_s \, {\widetilde{e}}(\ctrans^{-s}X)| \weight^{-w_N(X)}
\nonumber\\
&& + 2 |\epsilon| \sum_Y |h(Y)| \sum_Z \sum_{j::Z}
|e(Z) -{\widetilde{e}}(Z)|  \weight^{-w_N(Z \d (Y+j) \d \{j,j+1\})}
\nonumber\\
&& + 2 \sum_Y |h(Y)| \sum_Z \sum_{j::Z}
|e(Z) -{\widetilde{e}}(Z)|  \weight^{-w_N(Z \d (Y+j))}\nonumber\\
&& =: (a1) + (a2) + (a3) + (a4).
\label{boundaf}
\eea
We proved the following inequalities [\reff{wtria} and \reff{wtrib}]
in the previous section
\be
w_N((Y+j) \d Z) \le w_N(Y)+w_N(Z), \quad {\rm for} \quad j::Z
\label{wtriaq}
\ee
\be
w_N(X) \le w_N(\ctrans^{-s} X) + w_N(\ctrans^s \emptyset).
\label{wtribq}
\ee
In addition, we need the following two inequalities. 
\be
w_N((Y+j) \d Z \d \{j, j+1\}) \le w_N(Y)+w_N(Z) + 1, \quad {\rm for} 
\quad j::Z
\label{wtric}
\ee
\be
w_N(Z \d \{j, j+1\})\le w_N(Z) +1.
\label{wtrid}
\ee
Inequality \reff{wtric} can be proved with two applications
of \reff{wtriaq} as follows.
\bea
&& w_N((Y+j) \d Z \d \{j, j+1\}) = w_N([(Y \d \{N,1\})+j] \d Z)
\nonumber\\
&& \le w_N(Y \d \{N,1\}) + w_N(Z) \le w_N(Y)+ 1 + w_N(Z)
\eea
Similarly, inequality \reff{wtrid} follows from \reff{wtriaq}.

Using inequality \reff{wtrid} and eq.\reff{nz} we obtain
\bea
(a1) &\le& 2 |\epsilon| \sum_Z |e(Z) - {\widetilde{e}}(Z)|\left(|\epsilon|
  M\right)^{-w_N(Z)}\, \left(|\epsilon| M\right)^{-1}\,
\sum_{j\atop{j::Z}} 1,\nonumber\\
&=& 2 M^{-1}  \sum_{Z\atop{n(Z) \ne 0}} (n(Z) + 1)|e(Z) - {\widetilde{e}}(Z)| 
\left(|\epsilon| M\right)^{-w_N(Z)}.
\eea
We have added the
constraint $n(Z) \ne 0$ on the sum because $|e(Z) - {\widetilde{e}}(Z)|=0$ for
$n(Z) = 0$. Hence we can bound $(n(Z) + 1)$ in the above sum by
$2n(Z)$. This yields 
\be
(a1) \le 2M^{-1} ||e - {\widetilde{e}}||.
\label{a11}
\ee

Using the triangle inequality, 
\bea
|e_s \, e(\ctrans^{-s}X) - {\widetilde{e}}_s \, {\widetilde{e}}(\ctrans^{-s}X)| 
&\le&  
|e_s| |e(\ctrans^{-s}X) - \, {\widetilde{e}}(\ctrans^{-s}X)| 
+ |e_s - {\widetilde{e}}_s| |{\widetilde{e}}(\ctrans^{-s}X)|, 
\nonumber\eea
and \reff{wtribq} we get
\bea
(a2) 
&\le& ||e||\,\frac{1}{2} ||e-{\widetilde{e}}|| + ||e-{\widetilde{e}}|| \frac{1}{2} ||{\widetilde{e}}||.
\label{a22} 
\eea
Using \reff{wtric} and \reff{nz} we get
\bea
(a3) &\le& 
2 |\epsilon|\, \weight^{-1}\, \sum_Y |h(Y)| \weight^{-w_N(Y)}  
\sum_Z (n(Z) + 1) |e(Z) -{\widetilde{e}}(Z)|  \weight^{-w_N(Z)}
\nonumber\\
&\le& 2M^{-1} \bigl[ \sum_Y |h(Y)| \weight^{-w_N(Y)}\bigr]
 \, ||e - {\widetilde{e}}||.\nonumber\\
&\le&  2M^{-1}\bigl(e^{\delta} -1\bigr) \,||e - {\widetilde{e}}||, 
\label{a33}
\eea
where we have used \reff{wtrinb} - \reff{wtrinbcont}.

Similarly we get
\bea
(a4) &\le& 2 \sum_Y |h(Y)| \weight^{-w_N(Y)}\,\sum_Z (n(Z) + 1)
|e(Z) -{\widetilde{e}}(Z)|  \weight^{-w_N(Z)} \nonumber\\
&\le& 2 \bigl[\sum_Y |h(Y)| \weight^{-w_N(Y)}\bigr] \, ||e - {\widetilde{e}}||.\nonumber\\
&\le& 2 \bigl(e^{\delta} -1\bigr)\,  ||e - {\widetilde{e}}||.
\label{a44}
\eea
{From} \reff{a11}, \reff{a22}, \reff{a33} and \reff{a44} we obtain
\be
||F(e) - F({\widetilde{e}})|| \le K ||e - {\widetilde{e}}||,
\ee
where
\bea
K &=& \delta' + (1 +2M^{-1} ) (e^\delta - 1) + 2M^{-1}\nonumber\\
&=& \delta' + (1 + \frac{\delta}{2} ) (e^\delta - 1) + 
\frac{\delta}{2},
\eea
since we have chosen $\delta = 4M^{-1}.$
Hence, if $\delta'$ and $\delta$ are small enough then $K \le 1/2$.
To prove \reff{exboundtwo} we use \reff{emaf} and 
\reff{exaf} to compute
$\fp(0)$. Note that $e=0$ means that $e_s=0$ for all $s \in \L$,
and $e(X)=0$ for all $X$ except
$X=\emptyset $. We always have $e(\emptyset )=1$.
Letting ${\widetilde{e}}$ denote $F(0)$, we have 
\be
{\widetilde{e}}_m =  2\epsilon \sum_Y  h(Y) 1\bigl( Y \d \{N, 1\} = \ctrans^m 
\emptyset \bigr) - \sum_Y  h(Y) 1\bigl( Y = \ctrans^m 
\emptyset \bigr).
\label{abvaf}
\ee
For $X$ with $n(X) \ne 0$ 
\be 
{\widetilde{e}}(X) =  { 1 \over 2n(X)} 
 \, \Bigl[ 2\epsilon 1\bigl( X =\{N, 1\}\bigr) - 2 h(X) + 2 \epsilon 
h (X \d \{N, 1\})\Bigr].
\ee
Thus
\be
||\fp(0)|| \le 2 M^{-1} + 2 (e^{\delta} -1 ) + 2M^{-1} (e^\delta -1).
\ee
If we decrease $\delta$, then $K$ decreases. So we can assume that
$\delta$ is small enough that
$ 2(e^\delta-1) + {\delta e^\delta}/{2} < \delta'/2 $. So
\be
||\fp(e)|| \le ||\fp(e)-\fp(0)|| + ||\fp(0)||
\le {1 \over 2} \delta' + \frac{\delta}{2} + 2(e^\delta-1) + 
\frac{\delta}{2} (e^\delta-1)  \le \delta'
\ee
since $||e|| \le \delta'$.

This finishes the proof that the fixed point equation \reff{fpeaf} has 
a solution and thus completes the construction of eigenstates 
of $H$ with generalized momentum $k$. 
When $\epsilon=0$ these states are the lowest eigenstates in the 
subspaces of generalized momentum $k$ for $k \ne 0$, and the next to 
lowest for $k=0$. 
The same argument that we used in 
Section \ref{groundxz} proves that this is true for small $\epsilon$.
As in Section \ref{intfxz}, we can explicitly compute the lowest order
term in the dispersion relation for the interface and see that it is 
not zero. So the dispersion relation depends on $k$, indicating that the 
ground state does not correspond to a stable interface. 

\section{Ferromagnetic XXZ Chain}
\label{ferroxxz}
In this section we will prove that the ground state of the 
ferromagnetic chain has a stable interface at zero temperature
by showing that, for $s \ne 0$, 
the Fourier coefficients $e_s^N$ for the dispersion relation
vanish in the limit $N \rightarrow \infty$. 
Thus, in the infinite length limit the dispersion relation is flat, 
i.e., independent of the generalized momentum $k$. 
As discussed in the Introduction, the zero--temperature stability 
of the interface for the ferromagnet has been proven before.
The point of this section is to show that this result can also be 
obtained by our methods.
We will construct the wave function for ground states with 
an interface in them just as we did for the antiferromagnet. However,
we will use very different weights in the norm. The weight for 
the terms $e_s^N$ will be exponentially large in $N$ for $s \ne 0$. 
So the existence of a fixed point in this norm will prove that $e_s^N$ 
goes to zero exponentially fast as $N$ goes to infinity. 
The weights we use for the norm are based on considerations of 
how many applications of terms in the Hamiltonian it takes to get 
between various states. So we begin by studying the action of
the Hamiltonian.

A ferromagnetic XXZ chain of $N$ sites is governed by the 
Hamiltonian
\be
{\widetilde{H}} = - \sum_{j=1}^N \s^z_j \s^z_{j+1} 
- \epsilon \sum_{j=1}^N \s^x_j \s^x_{j+1}(1 - \s^z_j \s^z_{j+1}) 
\label{orfer}
\ee
(which is the ferromagnetic analog of \reff{oraf}). However,
unlike the antiferromagnetic case, we cannot force an interface into 
such a chain by considering $N$ to be odd and imposing periodic boundary 
conditions. So instead, to induce an interface we change the
coupling between the sites $N$ and $1$ as follows: We write the Hamiltonian
in the form
\be
\widetilde{H} = - \sum_{j=1}^N J_j \s^z_j \s^z_{j+1} 
- \epsilon \sum_{j=1}^N \s^x_j \s^x_{j+1} (1 - 
J_j \s^z_j \s^z_{j+1}).\nonumber\\
\label{orf}
\ee
If $J_j=1$ for all $j$ then \reff{orf} reduces to \reff{orfer}.
Such a Hamiltonian has two translation--invariant ground states 
-- with all spins up and all spins down, respectively. However,
the choice $J_N=-1$ and $J_j=1$ for all $j \ne N$, induces an interface
into the chain by causing at least one bond in the
chain to be frustrated. 
Moreover, this particular choice of 
coupling yields a unitarily equivalent Hamiltonian
$H$ [\reff{hxxzferro} below] which commutes with the generalized 
translation operator $\ctrans$ [\reff{genT}]. 
Hence, it allows us to exploit this symmetry to study the interface states, 
as in the case of the antiferromagnetic chain.

As before, we take $\fun$ to be the rotation operator defined by
\reff{rotop}. Hence,
\be
H := \fun \widetilde{H} \fun^{-1} = - \sum_{j=1}^N J_j \s^x_j \s^x_{j+1} 
- \epsilon \sum_{j=1}^N \s^z_j \s^z_{j+1} (1 - J_j  \s^x_j \s^x_{j+1}).
\label{hxxzferro}
\ee
The original Hamiltonian $\widetilde{H}$ [\reff{orf}] 
is not translation invariant 
when $J_N=-1$. Nonetheless, our choice of boundary conditions for the 
original Hamiltonian is such that the transformed Hamiltonian 
$H$ [\reff{hxxzferro}] commutes with $\ctrans$, as is easily checked. 
(Note that for the ferromagnetic chain we do not use the unitary transformation
$\sun$.)

Recall that ${\cal H}_{\Lambda}= (\C^2)^{\otimes |\Lambda|}$
is the Hilbert space of the lattice. 
In \reff{hxxzferro} the indices should be taken to be 
periodic e.g., $\sigma^x_{N+1}$ means 
$\sigma^x_1$. We can write the Hamiltonian as
\be
H = H_{0} + \hone,
\ee
where
\be
H_0 :=  - \sum_{j=1}^N J_j \s^x_j \s^x_{j+1},
\label{hoo}
\ee
and
\be
\hone:=  - \epsilon \sum_{j=1}^N \s^z_j \s^z_{j+1}
 (1 - J_j  \s^x_j \s^x_{j+1}). 
\ee
For any $X \subset \Lambda$ let $|X\rangle \in {\cal{H}}_\Lambda$
be given by
\be
|X\rangle = \sum_{\sigma} \sigma(X) |\sigma \rangle,
\label{ex3}
\ee
where $\sigma(X) = \prod_{j\in X} \sigma_j$. Hence,
\be
\sigma^x_i |X \rangle = \sum_\sigma \sigma(X) |\sigma^{(i)} \rangle,
\ee
where $\sigma^{(i)}$ is the spin configuration $\sigma$ but with 
$\sigma_i$ replaced by $-\sigma_i$. By making a change of variables in
the sum we obtain
\be
\sigma^x_i |X \rangle = \sum_\sigma \sigma^{(i)}(X) |\sigma\rangle,
\ee
where
\bea
\sigma^{(i)}(X) = - \sigma(X) \quad &{\hbox{if}}& \quad i \in X,\nonumber\\
 = \sigma(X) \quad &{\hbox{if}}& \quad i \not\in X.
\eea
Hence,
\bea
\sigma^x_i |X \rangle &= - & |X \rangle \quad {\hbox{for}} 
\quad i \in X,\nonumber\\
&=& |X \rangle \quad {\hbox{for}} \quad i \not\in X.
\eea
Note that the states $|X \rangle$ are eigenstates of $H_0$ [\reff{hoo}].

The ground states of the original Hamiltonian ${\widetilde{H}}$ [\reff{orf}]
for the choice $J_N =-1$ and $\eps = 0$ corresponds to a configuration 
consisting of a string of up--spins $(+)$ next to a string of 
down--spins $(-)$. We refer to such ground states of  ${\widetilde{H}}$
as its $\eps = 0$ interface states. However, the configuration
corresponding to the $\eps = 0$ interface states of the unitarily
equivalent Hamiltonian $H$ [\reff{hxxzferro}] (i.e., ground states
of $H_0$ [\reff{hoo}]) cannot be visualized as clearly. The 
unitary transformation $\fun$ obscures the picture. 
So, to describe $|X \rangle$, it is useful to think about 
$\fun^{-1} |X \rangle$. For example the state $\fun^{-1}|\emptyset\rangle$ 
corresponds to the configuration 
$$+ + + + + + \cdots + + +,$$
where the labels of the sites increase from $1$ to $N$ from left to right. 
It has a single interface between the nearest neighbor sites
$N$ and $1$. We say that there is an interface between two
nearest neighbor sites $j$ and $j+1$ if the nearest neighbor
bond $\langle j, j+1\rangle $ is frustrated, i.e., if the spins are antiparallel 
for $j \ne N$ and parallel for $j = N$. 
Note that for $X=\Lambda$, $\fun^{-1} |X \rangle$ is the configuration with
all $-$'s, this being another configuration with an interface between 
the sites $N$ and $1$. 
If $X=\{1,2, \cdots, j\}$, then $\fun^{-1} |X \rangle$ looks like 
$$+ + + \cdots + + - - \cdots - - -,$$
where the last $+$ occurs at the site $j$ ; 
for $X=\{j, j+ 1, \cdots, N\}$, $\fun^{-1} |X \rangle$ looks like 
$$- - - \cdots - + \cdots + + +, $$
where the first $+$ occurs at the site $j$.

\medskip

Let ${\cal{I}}(X)$ denote the set of sites for which the configuration 
corresponding to the state $\fun^{-1}|X\rangle$ has interfaces between each
site $i$ in this set and its nearest neighbor $i+1$. For 
$j \ne N$, $j \in {\cal{I}}(X)$ if and only if exactly one of $j$ and 
$j+1$ is in $X$, and for $j = N$, $j \in {\cal{I}}(X)$ if and only if 
both $N$ and $1$ are either in $X$ or outside it.
Then 
\be
\hone |X\rangle=  - 2 \epsilon \sum_{j \in {\cal{I}}(X)}
|X \d \{j, j+1\}\rangle.
\ee
Let $X,Y \subset \Lambda$ and consider the states 
$|X \rangle$ and $|Y \rangle$ .
If $|X|$ and $|Y|$ are both even (or both odd), 
then after repeated applications of the Hamiltonian on the state $|X\rangle$
we can obtain a state which has a nonzero overlap with $|Y\rangle$.
This is, however, not
possible if one of $|X|$ and $|Y|$ is even and the other is odd.
For $|X|$ and $|Y|$ both even (or odd) 
we define $\alpha(X \rightarrow Y)$ to be the 
minimum number of applications of the Hamiltonian necessary to get from 
$|X\rangle$ to a state which has a nonzero overlap with $|Y\rangle$.
We denote such a transition by the symbol $X \rightarrow Y$.
Hence, $\alpha(X \rightarrow Y)$ is equal to the smallest integer
$n$ for which
\be
\langle Y| \hone^n |X\rangle \ne 0 
\ee
Equivalently, we consider all sequences $X_0,X_1,X_2,\dots,X_n$
such that $X_0=X$, $X_n= Y$, and for each $k$ there is a 
$j:X_{k-1}$ so that  $X_k=X_{k-1} \d \{j,j+1\}$. Then $\alpha(X\rightarrow Y)$ 
is the smallest $n$ for all such sequences.  In addition, we define  
$\alpha(X):=\alpha(X \rightarrow \emptyset)$.
It is clear that $\alpha(X)$ is infinite for $|X|$ odd. 
Since $\ctrans^s$ is a unitary operator which commutes with 
$\hone$, we have
\be
\langle\ctrans^s Y| \hone^n |\ctrans^s X\rangle =
\langle Y| \ctrans^{-s} \hone^n \ctrans^s | X\rangle =
\langle Y| \hone^n | X\rangle 
\ee
The above equation implies that
\be 
\alpha(\ctrans^s X \rightarrow \ctrans^s Y) = 
\alpha(X \rightarrow Y).
\label{al1}
\ee

We start with the analog of eq.\reff{intffferro} for the 
case of Hamiltonian $H$ [\reff{hxxzferro}] (which is unitarily 
equivalent to the ferromagnetic Hamiltonian $\widetilde{H}$
\reff{orf}).  The change of the Hamiltonian $\widetilde{H}$
from the antiferromagnetic \reff{oraf} to the ferromagnetic \reff{orf} case 
(and hence the corresponding change of $H$ from  
\reff{hxxz} to \reff{hxxzferro})  changes some of 
the signs in eq.\reff{intffferro}. Moreover, since $h(X)=0$
for the Hamiltonian given by \reff{hxxzferro}, many of the terms 
in this equation reduce to zero. Taking into account these changes,
we obtain the following equation:
\bea
&&
 2 \sum_X n(X) \, e(X) \s (X)
 - 2 \epsilon \s_N \s_1 \, \sum_X e(X) \s (X) 
 \nonumber\\
&& 
- 2 \epsilon \sum_{j=1}^N \s_j \s_{j+1} \, s(j,N) \, 
\sum_{X :j} e(X) \s (X)
- \sum_{s=1}^{2N} \s_1 \cdots \s_s \, e_s 
\sum_X e(X) \s(X+s) = 0.
\nonumber\\ 
\label{intffferro2}
\eea

Using eqs. \reff{sjn} and \reff{tsx}, and picking out the 
coefficient of $\s(X)$ we have 
\bea
2 n(X) \, e(X) 
 &-& 2 \epsilon e(X \d \{N,1\}) 
 +  2 \epsilon e(X \d \{N,1\}) {\mathbf{1}}(N:X) \nonumber\\
&& 
 - 2 \epsilon \sum_{j=1}^{N-1} e(X \d \{j,j+1\}) {\mathbf{1}}(j:X) 
- \sum_{s=1}^{2N} e_s e(\ctrans^{-s} X) \quad \, = \quad 0,
\nonumber\\ 
\label{two1}
\eea
Since $1-{\mathbf{1}}(N:X)=
{\mathbf{1}}(N::X) \equiv {\mathbf{1}}(N \in {\cal{I}}(X))$, 
and for $j \ne N$, 
${\mathbf{1}}(j:X)={\mathbf{1}}(j \in {\cal{I}}(X))$, the above equation can be written as 
\be
2 n(X) \, e(X) 
- 2 \epsilon \sum_{j \in {\cal{I}}(X)} e(X \d \{j,j+1\}) 
- \sum_{s=1}^{2N} e_s e(\ctrans^{-s} X) \quad \, = \quad 0,
\ee
which we can write as 
\be
2 n(X) \, e(X) 
- 2 \epsilon \sum_{j=1}^{N} \sum_{Z\atop{ {\cal{I}}(Z)\ni j}} e(Z) 
1(Z \d \{j,j+1\}=X) 
- \sum_{s=1}^{2N} e_s e(\ctrans^{-s} X) \, = \, 0,
\nonumber\\
\label{intfgferro} 
\ee
since $j\in  {\cal{I}}(X)$ implies that $j \in {\cal{I}}(Z)$ 
for $X=Z \d \{j, j+1\}$.
\smallskip

\noindent
For $X$ such that $n(X) \ne 0$ we rewrite this as 
\be
e(X) = \frac{1}{2n(X)} \Bigl[
+ 2 \epsilon \sum_{j=1}^{N} \sum_{Z\atop{ {\cal{I}}(Z)\ni j}}
 e(Z) {\mathbf{1}}(Z \d \{j,j+1\}=X) 
+ \sum_{s=1}^{2N} e_s e(\ctrans^{-s} X) \Bigr].
\nonumber\\ 
\label{exferro}
\ee
\smallskip

\noindent
Recall that $n(X)=0$ if and only if $X$ is of the form $\ctrans^m(\emptyset)$
for some $m$. We assume that
$e(X) =0$ for all $X \subset \Lambda$ for which
$n(X) =0$, except for $X= \emptyset$ for which we assume
that
\be
e(\emptyset)=1.
\label{even}
\ee 
Hence, for $X=\ctrans^m(\emptyset)$, \reff{two1} becomes 
\be
e_m = 
 -2 \epsilon \sum_{j\in {\cal{I}}(X)} e(X \d \{j,j+1\}).
\nonumber\\ 
\label{em1}
\ee
When $X=\ctrans^m(\emptyset)$ the set ${\cal{I}}(X)$ contains only 
one site and we find that 
\be
e_m = -2 \epsilon e(X_m) \quad {\hbox{for}} \quad m \le N,
\label{em2}
\ee
where $X_m = \{1,2, \ldots m\} \d\{m, m+1\}$, and
\be
e_{m+N} = -2 \epsilon e(\Lambda \setminus X_m) =
 -2 \epsilon e(\ctrans^N X_m)  \quad {\hbox{for}} \quad m \le N\,.
\label{em3}
\ee
Note that $n(X_m) \ne 0$.
\smallskip

Consider the set of variables 
\be e := \{e(X): n(X) \ne 0\} \cup \{e_s: s=1,2,\cdots,2N \}
\ee
Equations \reff{exferro},\reff{em2} and \reff{em3} form a fixed point equation 
for $e$:
\be
\fp(e)=e
\label{mapferro}
\ee
Let us introduce the norm 
\be
||e|| := \sum_{m=1}^{2N} |e_m| (|\epsilon| M)^ {-\beta_m} + 
2 \sum_{X \atop{n(X) \ne 0}} |e(X)| n(X) (|\epsilon| M)^{-\alpha(X)},
\label{normferro}
\ee
where $M$ is a positive constant and 
$\beta_m = \alpha(\ctrans^m\emptyset \rightarrow \emptyset).$
Recall that $\alpha(X\rightarrow Y)$
is the least number of applications
of the Hamiltonian it takes to get from 
$|X\rangle$ to a state which has a nonzero overlap with $|Y\rangle$.
For $m$ odd, repeated applications of the Hamiltonian to 
$|\ctrans^m\emptyset\rangle$ can never produce a state with a nonzero 
overlap with $|\emptyset\rangle$.
So $\beta_m$ is taken to be infinite for odd values of $m$.
The factor of $2$ in the second term on the RHS of \reff{normferro}
is included merely for convenience. 
\smallskip

\noindent
For $m \le N$,
\bea
\beta_m &=& \alpha(\ctrans^m\emptyset)\nonumber\\
\beta_{m+N} &=&  \alpha(\ctrans^{m+N}\emptyset) \equiv  
\alpha(\Lambda \setminus \ctrans^m\emptyset). 
\eea
\smallskip

\noindent
\begin{theorem}
There exists a constant $M>0$ such that if $|\eps| M \le 1$,
then the fixed point equation \reff{mapferro} has a solution $e$, 
and $||e|| \le c $ for some constant $c$ which depends only on M.
Furthermore,
\be 
\sum_{s=-N+1, s \ne 0}^N |e_s| \le c (|\epsilon| M)^{N-1}
\label{fourier_bound}
\ee
So in the infinite length limit, the dispersion relation for an interface
is independent of the generalized momentum $k$. 
\label{ferro_thm}
\end{theorem}

\no {\bf Proof:} It is not hard to show that $\beta_2=\beta_{N-2}=N-1$,
and $\beta_s$ for other nonzero $s$ is even larger. So 
\reff{fourier_bound} will follow from the existence of a fixed point 
in the norm \reff{normferro}.
As before we prove the existence of a fixed point by proving 
\be
||\fp(e)-\fp({\widetilde{e}})|| \le \frac{1}{2}||e - {\widetilde{e}}|| \quad {\hbox{for}} \quad
||e||, ||{\widetilde{e}}|| \le \delta ;
\label{exboundferro}
\ee
\be
||\fp(e)|| \le \delta \quad {\hbox{for}} \quad
||e|| \le \delta. 
\label{exboundferro2}
\ee
with 
\be
\delta = {4 \over M}
\ee

\smallskip

\noindent
{From} \reff{em2} and \reff{em3} we get (using the definition of $\beta_m$)
\bea
\sum_{m=1}^{2N} |e_m -{\widetilde{e}}_m| (|\epsilon| M)^{-\beta_m} &\le&
\sum_{m=1}^{N} |e_m -{\widetilde{e}}_m| (|\epsilon| M)^{-\beta_m} +
\sum_{m=1}^{N} |e_{m+N} -{\widetilde{e}}_{m+N}| (|\epsilon| M)^{-\beta_{m+N}}\nonumber\\
 &=& 2|\epsilon| \, (|\epsilon| M)^{-1}\sum_{m=1}^{N} |e(X_m) -{\widetilde{e}}(X_m)| (|\epsilon| M)^{-\alpha(X_m)} 
\nonumber\\
\quad \, &+&
2|\epsilon| \,(|\epsilon|M)^{-1}\sum_{m=1}^{N} |e(\Lambda \setminus X_m) - 
{\widetilde{e}}(\Lambda \setminus X_m)|
(|\epsilon| M)^{-\alpha(\Lambda \setminus X_m)}.
\nonumber\\
\label{bd1}
\eea
This is because, for $m \le N$, 
$$
\beta_m = \alpha(\ctrans^m \emptyset) = \alpha(X_m) + 1, $$ 
and
$$
\beta_{m+N} = \alpha(\ctrans^{N+m} \emptyset) = \alpha(\Lambda \setminus X_m) 
+ 1, $$ 
Hence, 
\bea
{\hbox{RHS of \reff{bd1}}} &\le& 2 |\epsilon|\,(|\epsilon|M)^{-1} \sum_{Y\atop{n(Y) \ne 0}}
|e(Y) - {\widetilde{e}}(Y)| (|\epsilon| M)^{-\alpha (Y)}\nonumber\\
&\le&
M^{-1}\, ||e - {\widetilde{e}}||. 
\eea
Further, from \reff{exferro} we get
\bea
&&
2\sum_{X\atop{n(X) \ne 0}} n(X) |e(X) - {\widetilde{e}}(X)| (|\epsilon| M)^{-\alpha(X)}\nonumber\\
&& \le 
2|\epsilon| \sum_{X\atop{n(X) \ne 0}} \sum_{j=1}^{N}
\sum_{Z\atop{{\cal{I}}(Z) \ni j}} |e(Z) - {\widetilde{e}}(Z)| (|\epsilon| M)^{-\alpha(Z \d \{j, j+1\})}
1(Z \d \{j, j+1\}=X)
\nonumber\\
&& + \sum_s \sum_{X\atop{n(X) \ne 0}} |e_s e(\ctrans^{-s}X) - {\widetilde{e}}_s {\widetilde{e}}(\ctrans^{-s}X)| 
(|\epsilon| M)^{- \alpha(X)}.
\nonumber\\
&& =: (a) + (b).
\label{bdex}
\eea
We claim that 
\be
\alpha(Z \d \{j, j+1\}) \le 
\alpha(Z) + 1 \quad {\hbox{for}}\quad {j\in{\cal{I}}(Z) }.
\label{alpha2}
\ee
To see this note that if $j\in {\cal{I}}(Z)$, then 
$j\in {\cal{I}}(Z \d \{j,j+1\})$. So a single application of the 
Hamiltonian can cause the transition $Z \d \{j,j+1\} \rightarrow Z$.
Using \reff{alpha2} we get  
\bea
(a) &\le& 2|\epsilon|  \, (|\epsilon| M)^{-1}\,
\sum_{Z \atop{n(Z) \ne 0}} |e(Z) - {\widetilde{e}}(Z)| 
(|\epsilon| M)^{-\alpha(Z)} \sum_{j\atop{j\in {\cal{I}}(Z)}} 1\nonumber\\
&\le& 2 M^{-1}\, \sum_{Z \atop{n(Z) \ne 0}} 
|e(Z) - {\widetilde{e}}(Z)| 
(|\epsilon| M)^{-\alpha(Z)} |\delta Z|\nonumber\\
 &\le& 2 M^{-1}\, \sum_{Z \atop{n(Z) \ne 0}} 
|e(Z) - {\widetilde{e}}(Z)| 
(|\epsilon| M)^{-\alpha(Z)} (n(Z) + 2)\nonumber\\
 &\le& {3}\, M^{-1}\, ||e - {\widetilde{e}}||,
\label{a}
\eea
where we have used the inequality
$$ |\delta Z| \le n(Z) + 2.$$
\smallskip

\noindent
Moreover, using the triangle inequality we get
\be
(b) \le
\sum_{s=1}^{2N} \sum_{X \atop{n(X) \ne 0}} \Bigl[ 
|e_s| \, |e(\ctrans^{-s}X) - {\widetilde{e}}(\ctrans^{-s}X)| \, (|\epsilon| M)^{-\alpha(X)} 
+  |e_s-{\widetilde{e}}_s| \, |{\widetilde{e}}(\ctrans^{-s}X)| (|\epsilon| M)^{-\alpha(X)}\Bigr].
\label{c1}
\ee
Let $Y = \ctrans^{-s} X$. Hence, $X= \ctrans^s Y$. Since $n(X) \ne 0$ in the above
sum, we must have $Y \ne \emptyset$ and $n(Y) \ne \emptyset$. 
We claim that 
\be
\alpha( \ctrans^s Y ) \le \alpha(Y) + \beta_s.
\label{bbs}
\ee
If we have a sequence of applications of the Hamiltonian that causes
the transition $\ctrans^s Y \rightarrow \ctrans^s \emptyset$ 
and another sequence that causes the transition 
$\ctrans^s \emptyset \rightarrow \emptyset$, then together they
give a sequence which results in $\ctrans^s Y \rightarrow \emptyset$. Thus
\be
\alpha( \ctrans^s Y )= \alpha( \ctrans^s Y \rightarrow \emptyset)
\le \alpha( \ctrans^s Y \rightarrow \ctrans^s \emptyset) +
\alpha ( \ctrans^s \emptyset \rightarrow \emptyset) 
=\alpha(Y) + \alpha( \ctrans^s \emptyset) = \alpha(Y) + \beta_s
\label{ee23}
\ee
where we have used \reff{al1}.
{From} \reff{c1} and \reff{bbs} it follows that
\bea
(b) &\le&
\sum_{s=1}^{2N} |e_s|\, (|\epsilon| M)^{-\beta_s}  \sum_{Y \atop{n(Y) \ne 0}} \,
|e(Y) - {\widetilde{e}}(Y)|\,(|\epsilon| M)^{- \alpha(Y)}\nonumber\\
& & \quad +  \sum_{s=1}^{2N}|e_s-{\widetilde{e}}_s| \,(|\epsilon| M)^{-\beta_s} 
\sum_{Y \atop{n(Y) \ne 0}} \,
|{\widetilde{e}}(Y)|\,(|\epsilon| M)^{- \alpha(Y)}\label{tt}\\
 &\le& \frac{1}{2} ||e||\, ||e - {\widetilde{e}}|| +  \frac{1}{2} ||{\widetilde{e}}||\, ||e - {\widetilde{e}}||.
\label{b}
\eea
{From} \reff{a} and \reff{b} it follows that 
\bea
{\hbox{RHS of \reff{bdex}}} &\le& ||e - {\widetilde{e}}||\Bigl[  \frac{1}{2} ||e||
+  \frac{1}{2} ||{\widetilde{e}}|| + {4}M^{-1} \Bigr] 
\le K ||e - {\widetilde{e}}|| \nonumber \\
\label{bd2}
\eea
where we have used $||e|| \le \delta$, $||{\widetilde{e}}|| \le \delta$, and defined  
\bea
K &=& \delta + 4 M^{-1}= 2\delta,
\eea
If $\delta \le 1/4$, then $K \le 1/2$. 

To prove \reff{exboundferro2}, we use \reff{exferro}, \reff{em2} and 
\reff{em3} to compute
$\fp(0)$. Note that $e=0$ means that $e_s=0$ for all $s \in \L$,
and $e(X)=0$ for all $X$ except
$X=\emptyset $. We always have $e(\emptyset )=1$. 
Letting ${\widetilde{e}}$ denote $F(0)$, we have 
\be
{\widetilde{e}}_m = 0 \quad {\hbox{for all}} \quad m.  
\label{abv2}
\ee
For $X = \{N, 1\}$ we have
\be 
{\widetilde{e}}(X) =  { 1 \over 2n(X)} \bigl[  2 \epsilon \bigr]
\ee
and ${\widetilde{e}}(X) = 0$ for all other $X$ for which $n(X) \ne 0$.
Thus 
\be
||F(0)|| \le 2 |\epsilon| (|\epsilon| M)^{-\alpha(\{N, 1\})} = 
2|\epsilon|\,(|\epsilon| M)^{-1} = 2 M^{-1},
\ee
since $\alpha(\{N, 1\})=1$. 
Hence, for $||e|| \le \delta$, where $\delta = 4M^{-1}$
\be
||\fp(e)|| \le ||\fp(e)-\fp(0)|| + ||\fp(0)||
\le {1 \over 2} ||e|| + 2 M^{-1} \le \delta 
\ee
This finishes the proof that the fixed point equation has 
a solution and so completes the proof of Theorem \ref{ferro_thm}.

\section*{Acknowledgements}
We would like to thank B. Nachtergaele for helpful
suggestions. ND would also like to thank Y.M. Suhov
for interesting discussions. TK acknowledges the support of the 
National Science Foundation (DMS-9970608 and DMS-0201566).

\bigskip

\noindent

\bibliographystyle{amsalpha}

\end{document}